\documentclass{aa}
\usepackage{amsmath}
\usepackage{graphicx}
\usepackage{placeins}
\usepackage{float}
\usepackage{txfonts}
\usepackage{hyperref}
\usepackage[dvipsnames]{xcolor}

\usepackage[symbol]{footmisc}

\begin{document}

   \title{PGIR\,20eid (SN\,2020qmp): A Type IIP Supernova at 15.6 Mpc discovered by the Palomar Gattini-IR survey \thanks{Table 1 is available in electronic form at http://cdsweb.u-strasbg.fr/cgi-bin/qcat?J/A+A/}}
   \author{G. P. Srinivasaragavan \inst{1,2} \and I. Sfaradi \inst{3} \and J. Jencson \inst{4} \and K. De \inst{5} \and A. Horesh \inst{3} \and M. M. Kasliwal \inst{1} \and S. Tinyanont \inst{6} \and M. Hankins \inst{7} \and S. Schulze \inst{8} \and M. C. B. Ashley \inst{9} \and  M. J. Graham \inst{1} \and V. Karambelkar \inst{1} \and R. Lau \inst{10} \and A. A. Mahabal \inst{11, 12} \and A. M. Moore \inst{13} \and E. O. Ofek \inst{14} \and Y. Sharma \inst{1} \and J. Sollerman \inst{15} \and J. Soon \inst{13} \and R. Soria \inst{16, 17}\and T. Travouillon \inst{13} \and R. Walters \inst{18} 
          }
   \institute{Cahill Center for Astrophysics, California Institute of Technology, 1200 E. California Blvd, Pasadena, CA 91125, USA \and Department of Astronomy, University of Maryland, College Park, MD 20742, USA \and Racah Institute of Physics, The Hebrew University of Jerusalem, Jerusalem 91904, Israel \and Steward Observatory, University of Arizona, 933 North Cherry Avenue, Rm. N204, Tucson, AZ 85721-0065, USA  \and MIT-Kavli Institute for Astrophysics and Space Research, 77 Massachusetts Ave., Cambridge, MA 02139, USA \and Department of Astronomy and Astrophysics, University of California, Santa Cruz, CA 95064, USA \and The Department of Physical Sciences, Arkansas Tech Univesrsity, 215 West O Street, Russellville, AR 72801, USA \and The Oskar Klein Centre, Physics Department, Stockholm University, Albanova University Centre, SE 106 91 Stockholm, Sweden \and School of Physics, University of New South Wales, Sydney NSW 2052, Australia \and Institute of Space \& Astronautical Science, Japan Aerospace Exploration Agency, 3-1-1 Yoshinodai, Chuo-ku, Sagamihara, Kanagawa
252-5210, Japan \and Division of Physics, Mathematics and Astronomy, California Institute of Technology, Pasadena, CA 91125, USA \and Center for Data Driven Discovery, California Institute of Technology, Pasadena, CA 91125, USA \and Research School of Astronomy and Astrophysics, Australian National University, Canberra, ACT 2611, Australia \and Benoziyo Center for Astrophysics, Weizmann Institute of Science, 76100 Rehovot, Israel \and The Oskar Klein Centre, Department of Astronomy, Stockholm University,
Albanova, SE-106 91 Stockholm, Sweden \and College of Astronomy and Space Sciences, University of the Chinese Academy of Sciences, BeÄ³ing 100049, China \and Sydney Institute for Astronomy, School of Physics A28, The University of Sydney, Sydney, NSW 2006, Australia \and Caltech Optical Observatories, California Institute of Technology, Pasadena, CA 91125, USA }
             
   \date{}
  \abstract
   {}
 {We present a detailed analysis of SN 2020qmp, a nearby Type {IIP} core-collapse supernova (CCSN) that was discovered by the Palomar Gattini-IR survey in the galaxy UGC07125 (distance of $\approx 15.6 \pm 4$ Mpc). We illustrate how the multiwavelength study of this event helps our general understanding of stellar progenitors and circumstellar medium (CSM) interactions in CCSNe. We highlight the importance of near-infrared (NIR) surveys for detections of supernovae in dusty environments.}
     {We analyze data from observations in various bands: radio, NIR, optical, and X-rays.  We use optical and NIR data for a spectroscopic and spectro-polarimetric study of the supernova and to model its light curve (LC). We obtain an estimate of the zero-age main-sequence (ZAMS) progenitor mass from the luminosity of the [O I] doublet lines ($\lambda\lambda 6300,6364$) normalized to the decay power of $^{56}$Co. We also independently estimate the explosion energy and ZAMS progenitor mass through hydrodynamical LC modeling. From radio and X-ray observations, we derive the mass-loss rate and microphysical parameters of the progenitor star, and we investigate possible deviations from energy equipartition of magnetic fields and electrons in a standard CSM interaction model. Finally, we simulate a sample of CCSNe with plausible distributions of brightness and extinction, within 40 Mpc, and test what fraction of the sample is detectable at peak light by NIR surveys versus optical surveys.}
    {SN 2020qmp displays characteristic {hydrogen} lines in its optical spectra as well as a plateau in its optical LC, hallmarks of a Type {IIP} supernova. We do not detect linear polarization during the plateau phase, with a 3$\sigma$ upper limit of 0.78\%. Through hydrodynamical LC modeling and an analysis of its nebular spectra, we estimate a ZAMS progenitor mass of around {11.0 $\mathrm{M_\odot}$ and an explosion energy of around $0.8 \times 10^{51}$ erg.} We find that the spectral energy distribution cannot be explained by a simple CSM interaction model, assuming a constant shock velocity and a steady mass-loss rate. In particular, the excess X-ray luminosity compared with the synchrotron radio luminosity suggests deviations from equipartition. Finally, we demonstrate the advantages of NIR surveys over optical surveys for the detection of dust-obscured CCSNe in the local Universe. Specifically, our simulations show that the Wide-Field Infrared Transient Explorer will detect {up to} 14 more CCSNe (out of the 75 expected in its footprint) within 40 Mpc over five years than would an optical survey equivalent to the Zwicky Transient Facility. }
   {We have determined or constrained the main properties of SN 2020qmp and its progenitor, highlighting the value of multiwavelength follow-up observations of nearby CCSNe. {We have shown that forthcoming NIR surveys will enable us to improve constraints on the local CCSN rate by detecting obscured supernovae that would be missed by optical searches.}}

    \titlerunning{Discovery of SN 2020qmp}
    \authorrunning{G.P. Srinivasaragavan}
   \maketitle

\section{Introduction}
\label{introduction}
Type II supernovae (SNe) are hydrogen-rich core-collapse supernovae (CCSNe) that represent the fate of stars that have a minimum mass of around 7 to 9  $\mathrm{M}_{\odot}$ \citep{Smartt2009}, though the maximum mass of CCSN progenitors is a debated topic \citep{ Utrobin2009,Dessart2010,Jerkstrand2012}. The Type II class is divided observationally into many different subclasses based on their light curves (LCs) and spectroscopic properties, including types IIP, IIL, {IIn, and IIb} \citep{Filippenko1997,Gal-Yam2017}. Of these, Type IIP events, which are characterized by a plateau in their optical LCs that lasts for about 100 days after the explosion, are the most common \citep{Branch2017}. 

Though Type IIP SNe are among the most common SNe found, it is uncommon to discover nearby CCSNe (only five during the past three years within 10 Mpc have been reported to the Transient Name Server\footnote{https://www.wis-tns.org/}). Nearby and {bright} CCSNe allow us to probe many different facets of SN physics by, for example, obtaining high-resolution spectra for astrochemistry purposes \citep{Shivvers_2015}, astrometrically pinpointing the progenitor star \citep{ Smartt2009,Smartt2015}, analyzing the physics of the shock breakout \citep{Rabinak_2011, Sapir_2017}, understanding the polarimetry of the SN  \citep[eg.][]{Leonard2006, Wang2008, Nagao2017, Nagao2018, Tinyanont2019a}, and opening the avenue for multi-messenger follow-up on the sources \citep{Nakamura2016}. Furthermore, the interaction between the blast wave of CCSNe and the circumstellar medium (CSM) or interstellar medium generates {multiwavelength} emission through synchrotron radiation processes \citep{Chevalier1998}. Observing this synchrotron radiation, mainly in the radio and X-ray, provides key insights into the progenitor star's final years and allows us to probe the very late stages of the stellar evolution of massive stars \citep{ Berger2002, BenAmi12, Assaf2013}. 

Palomar Gattini-IR (PGIR; \citealt{Moore2020, de2020}) is a wide-field near-infrared (NIR) time-domain survey that is able to image three-fourths of the accessible night sky on a given night. Located at Palomar Observatory, PGIR {uses a telescope with an aperture} of 300 mm and a camera field of view of 25 square degrees, along with a HAWAII-2RG detector that operates in a single $J$-band filter \citep{de2020}. {PGIR has a median cadence of 2 days and can image sources up to} a median depth of $15.7$\,AB mag \citep{de2020} in the $J$ band. As a wide and shallow infrared time-domain survey, PGIR is sensitive to bright NIR transients in the Milky Way and nearby galaxies, including events that could be missed in the optical due to a large extinction.
\newline
\indent On July 30, 2020 (all dates are in UT time), PGIR made its first extragalactic discovery of a {SN} with its detection of PGIR 20eid (SN 2020qmp), which was  spectroscopically classified as a Type {IIP} SN \citep{ATEL}. In this paper, we present the NIR and optical LCs and spectroscopy of the SN up to the first 244 days of its evolution, {as well as polarimetry measurements from the infrared spectropolarimetery mode of the Wide-field Infrared Camera on the 200-inch Hale Telescope at Palomar Observatory (WIRC+Pol; \citealt{Tinyanont2019a})}. We then analyze the optical spectra of the SN after it has reached its nebular phase in order to infer the mass of the progenitor star. We also infer the mass of the progenitor star and the SN's explosion energy through hydrodynamical LC modeling. We also present radio data obtained from the Karl G. Jansky Very Large Array (VLA) and X-ray data from the {Neil Gehrels \textit{Swift} Observatory} \citep[\textit{Swift};][]{Gehrels_2004}, which allowed us to infer key characteristics pertaining to properties of the SN blast wave and its interaction with materials lost from the progenitor star during the late stages of its life. Finally, we comment on the local CCSN rate and on how NIR surveys are well equipped to find optically obscured CCSNe in the future due to their ability to see through large amounts of dust extinction.
\newline
\indent The paper is organized as follows. In Sect.  \ref{Observations} we present the observations {in the ultraviolet (UV), optical, and NIR by \textit{Swift}, the Zwicky Transient Facility \citep[ZTF;][]{Bellm2019,Graham2020, Masci2019, Dekaney2020}, and PGIR}, along with optical and NIR spectra and radio observations. In Sect.  \ref{Nebularsection} we analyze the nebular spectra of the SN and compare it with model spectra in order to  infer the zero-age main-sequence (ZAMS) progenitor mass. In Sect.  \ref{LCModeling} we present comparisons of hydrodynamical LC models to the observed LCs of the SN in order to again infer the ZAMS progenitor mass using an independent method and to infer the explosion energy of the SN. In Sect.  \ref{progenitor} we present the analysis from the radio and X-ray data and infer various properties of the blast wave and the progenitor star's mass-loss rate, as well as possible deviations from a standard CSM interaction model. In Sect. \ref{rate} we describe the local CCSN rate and examine the sensitivity of PGIR to highly extinguished SNe in the local Universe compared to optical searches. Finally, in Sect.  \ref{Conclusions} we summarize the main conclusions of our results. 
\begin{figure*}
    \centering
    \includegraphics[width=0.95\linewidth]{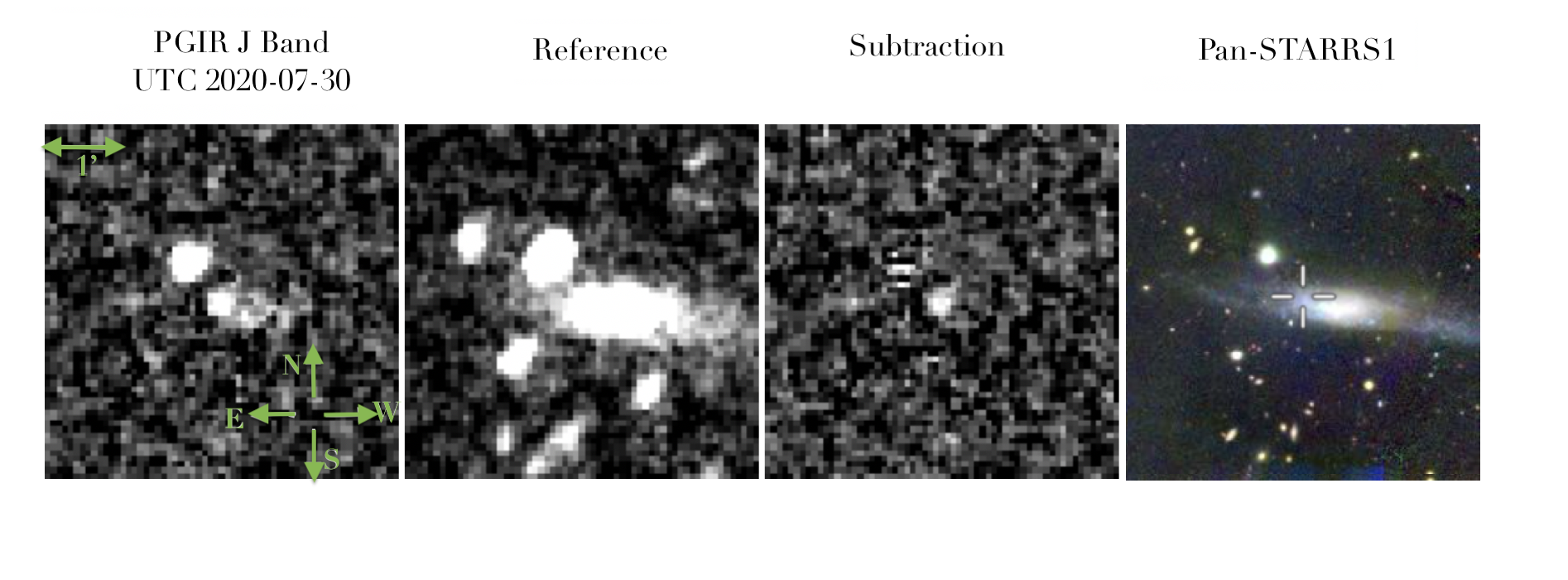}
    \caption{Discovery {location} of SN 2020qmp, containing images from the night of discovery (July 30, 2020) taken by PGIR. The left frame is the first detection, the next frame to the right is a reference template image constructed from stacking previous PGIR images, and the subtraction image clearly shows the source coincident with the host galaxy. A pre-explosion optical image from Pan-STARRS1 is shown for comparison, with the position of the SN in the crosshairs.}
    \label{DiscoveryFigure}
\end{figure*}
\begin{figure*}
\centering
    \includegraphics[width=0.85\linewidth]{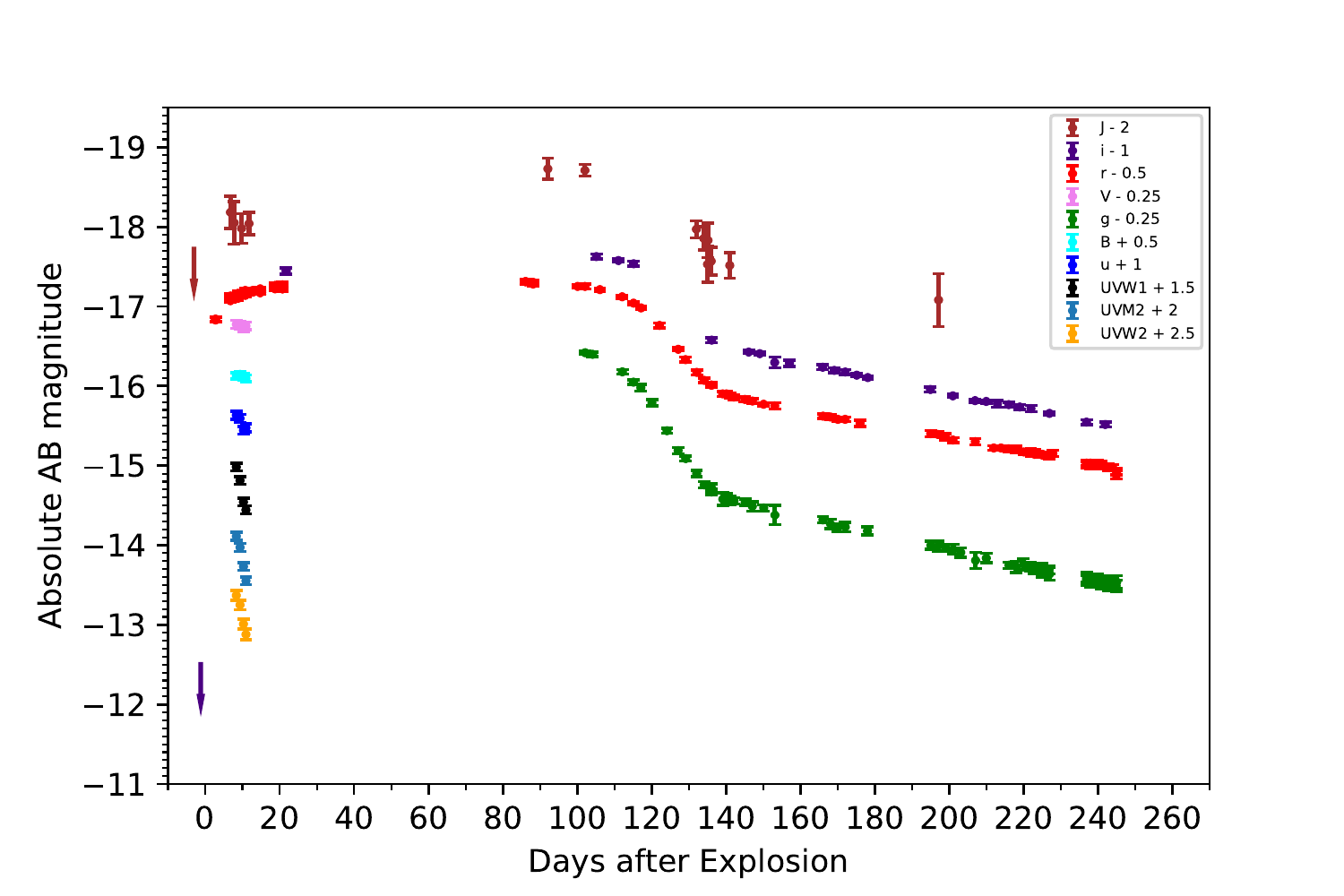}
    \caption{Light curve of SN 2020qmp. The LC includes photometry points from the PGIR survey {($J$ band)} as well as from the \textit{Swift} UVOT telescope ({$UVW1$, $UVM2$, $UVW2$, $u$, $B$, and $V$ bands)} {and ZTF ($g$, $r$, and $i$ bands)}.}
    \label{LightcurveSpectra}
\end{figure*}

\begin{figure}
    \centering
    \includegraphics[width = \linewidth]{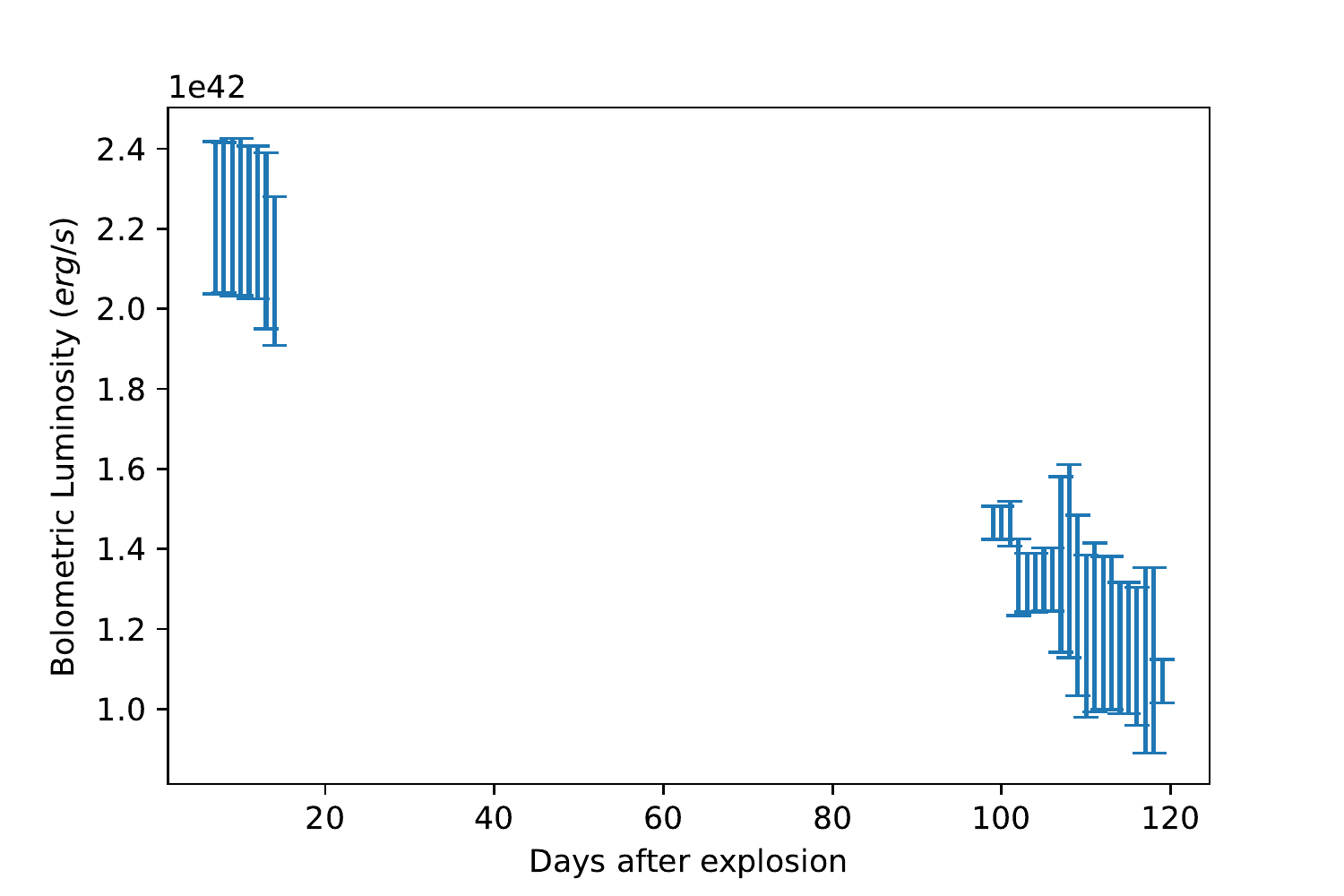}
    \includegraphics[width = \linewidth]{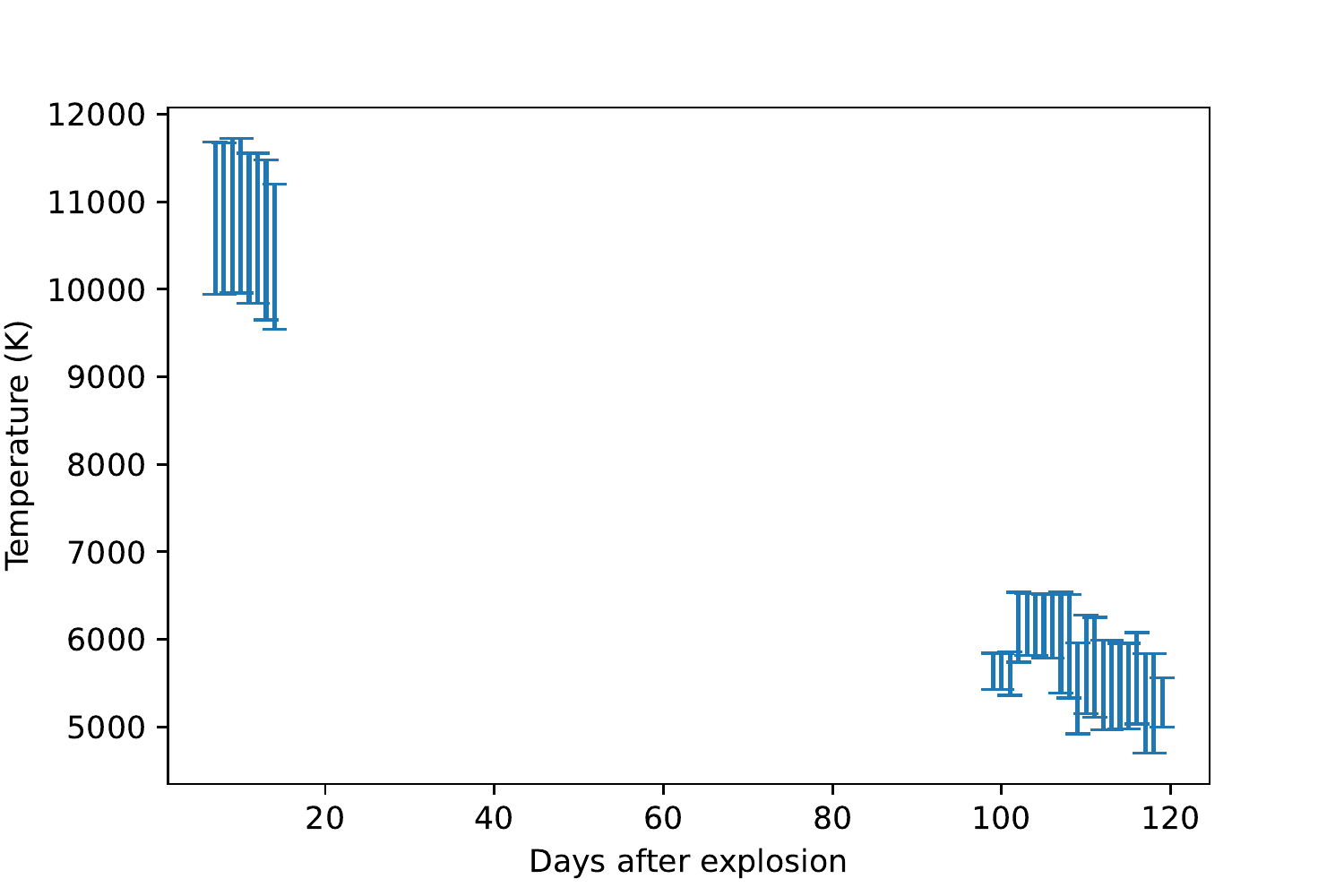}
    \caption{Bolometric LC and temperature evolution of SN 2020qmp until the end of the plateau phase. \textit{Top panel:} Bolometric LC of SN 2020qmp up {until the end of the plateau phase. The errors displayed in the plot are obtained from individual measurement errors, but we note that an overall distance uncertainty to the source of 0.54 mag dominates the overall error.} \textit{Bottom panel:} {Temperature evolution over the plateau phase of the blackbody fits used to derive the bolometric luminosities.} }
    \label{Bollc}
\end{figure}

\begin{figure*}
\centering
\includegraphics[width=0.75\linewidth]{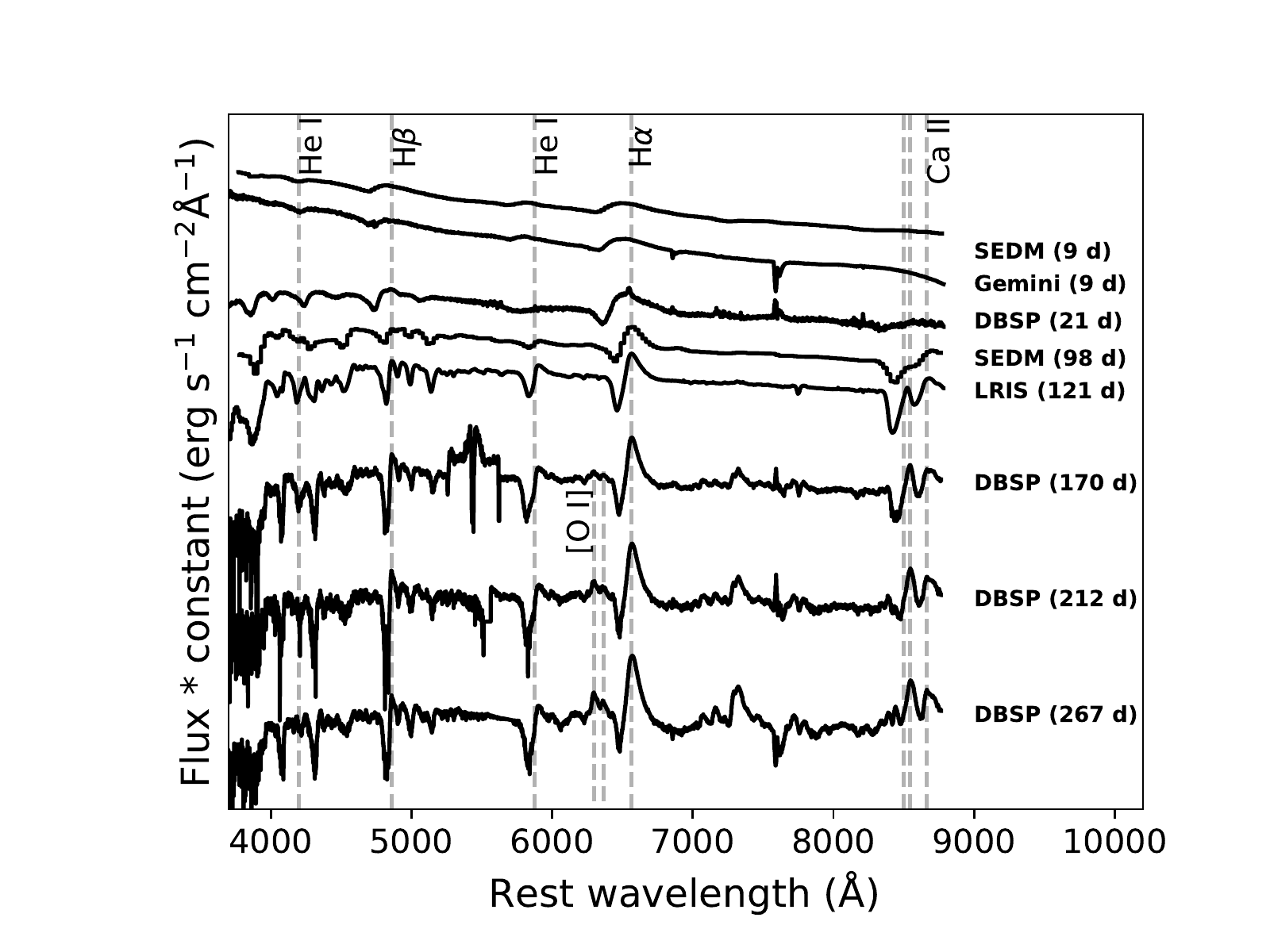}
 \includegraphics[width = 0.75\linewidth]{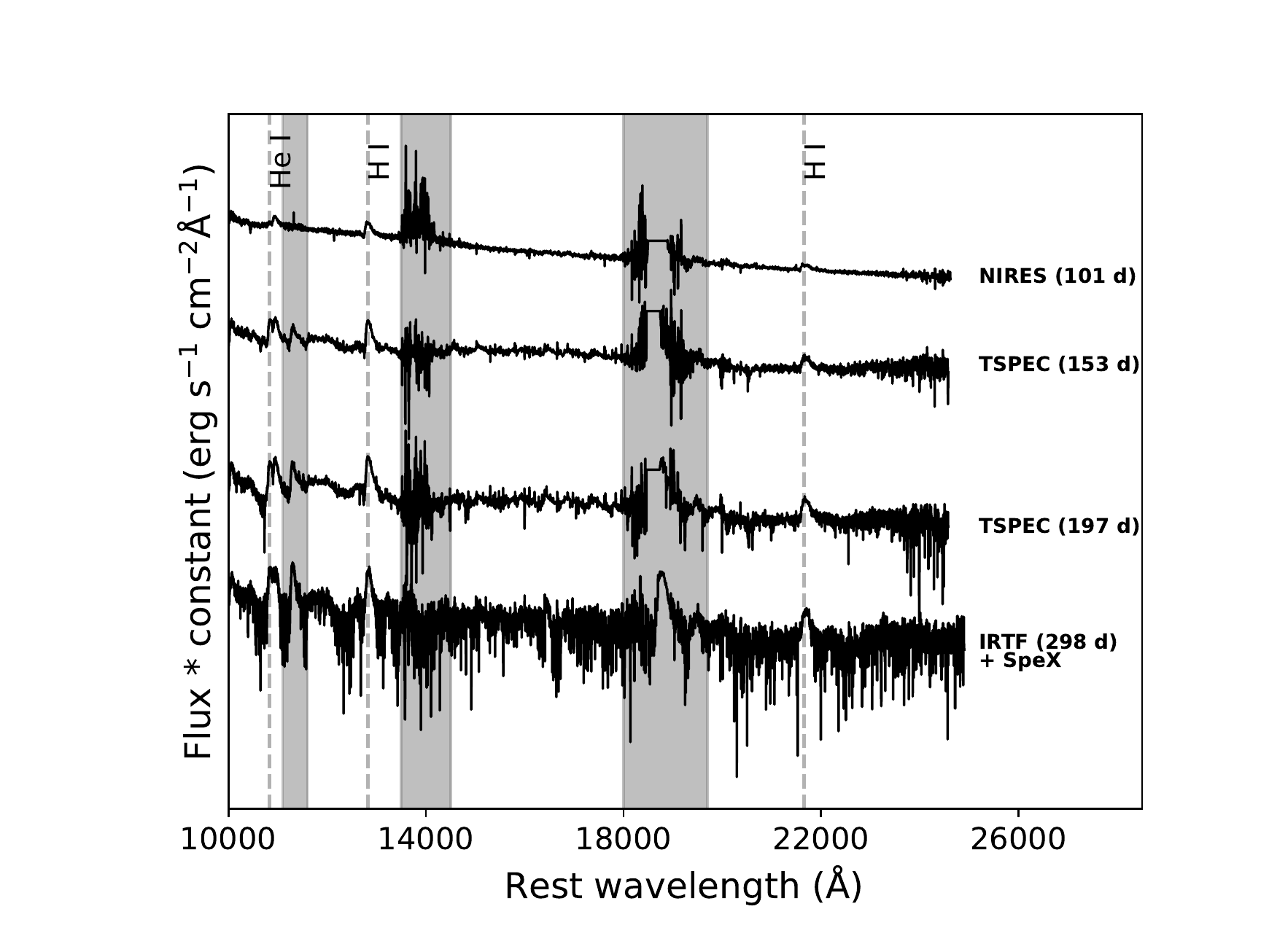}
\caption{Spectra obtained of SN 2020qmp, in the optical and NIR. \textit{Top panel}: Spectral evolution in the optical of SN 2020qmp. The most prominent spectral features are labeled, with the instruments and phases shown to the right of the spectrum. \textit{Bottom panel}: Spectral evolution in the NIR of SN 2020qmp, with the same labels {as in} the top panel. {Areas of atmospheric absorption are indicated with gray bands.} }
\label{Spectra}
\end{figure*}

\section{Observations of SN 2020qmp}
\label{Observations}
\subsection{Photometric and X-ray observations}
\label{photometric}
\indent SN 2020qmp was first discovered in the automated image subtraction and transient detection pipeline of the PGIR survey on July 30, 2020, at a (RA, Dec) of {($12^{\rm{h}}08^{\rm{m}}44.43^{\rm{s}}$}, +36:48:19.4) at magnitude $J = 14.74 \pm 0.2$\,AB mag (all magnitudes for the rest of the paper are in units of AB mag unless specified otherwise). {This source was detected as part of the search for large amplitude transients described in \citet{De2021}}. The transient was detected on the spiral arm of galaxy UGC07125. {There are eight different Tully-Fischer distances \citep{TullyFischerPaper} provided by the Nasa Extragalatic Database (NED). Using Tully's Cosmic Flow Calculator \citep{Shaya2017, Graziani2019}, the galactocentric {velocity} of UGC07125 is 1089 $\pm$ 2 km/s. This galactocentric speed makes distances of less than 10 Mpc and greater than 25 Mpc unlikely; five of the distances from NED are outside this 10 Mpc to 25 Mpc range. Furthermore, the Tully-Fischer distance derived from the B band \citep{Bbanddistance} provided by NED is more affected by extinction than those derived in the NIR. Therefore, out of the two Tully-Fischer distances left, we chose $15.6^{+4.4}_{-3.4}$ Mpc \citep{Tullydistance}, which is still is within the error range of the other distance, $17.5^{+3.9}_{-3.1}$ \citep{2014distance}. The distance we chose corresponds to a distance modulus of  $m - M = 30.97 \pm 0.54$ mag}. Assuming this distance, the absolute magnitude of the transient upon first detection was $M = -15.7$ in the $J$ band. The latest non-detection by PGIR was on July 25, 2020, up to a $5\sigma$ limiting magnitude of $J=14.8$\,mag; this is five days before the first detection due to the low visibility of the field, as it was close to the Sun. The discovery location of the SN by PGIR along with an image from The Panoramic Survey Telescope and Rapid Response System (Pan-STARRS1) survey \citep{Kaiser2002} are shown for reference in Fig. \ref{DiscoveryFigure}.

Though PGIR made the initial discovery of the SN, a search in ZTF data at the same position revealed an even earlier detection of the SN, on July 26, 2020 {\citep[ZTF20abotkfn;][]{ATEL}} at a magnitude of $r = 14.6$. The latest non-detection by ZTF was on July 22, 2020, up to a limiting magnitude of {$i=19.1$\, mag}. {The Gaia mission \citep{Gaia2016} reports a detection on July 24, 2020, at 13:46:33.6 of G = 14.9 mag. Therefore, we estimate the explosion date to be July  23, 2020, at 9:15:07.2, which is the average time between the latest non-detection by ZTF and the first detection made by Gaia.} All mentions of "days" used in figures are with reference to this explosion date. ZTF continued to observe the SN in the  $g$, $r$, and $i$ bands. {The ZTF photometry was retrieved from the ZTF transient detection pipeline. Transients in the difference imaging pipeline (based on the Zackay, Ofek, and Gal-Yam subtraction algorithm; \citealt{Zackay2016}) of ZTF are reported and distributed in Avro alert packets2
\citep{Patterson2019}, including photometry and metadata for the detected transient as well as a 30-day history for the previous detections and non-detections.}

\begin{figure}
    \centering
    \includegraphics[width = 0.9\linewidth]{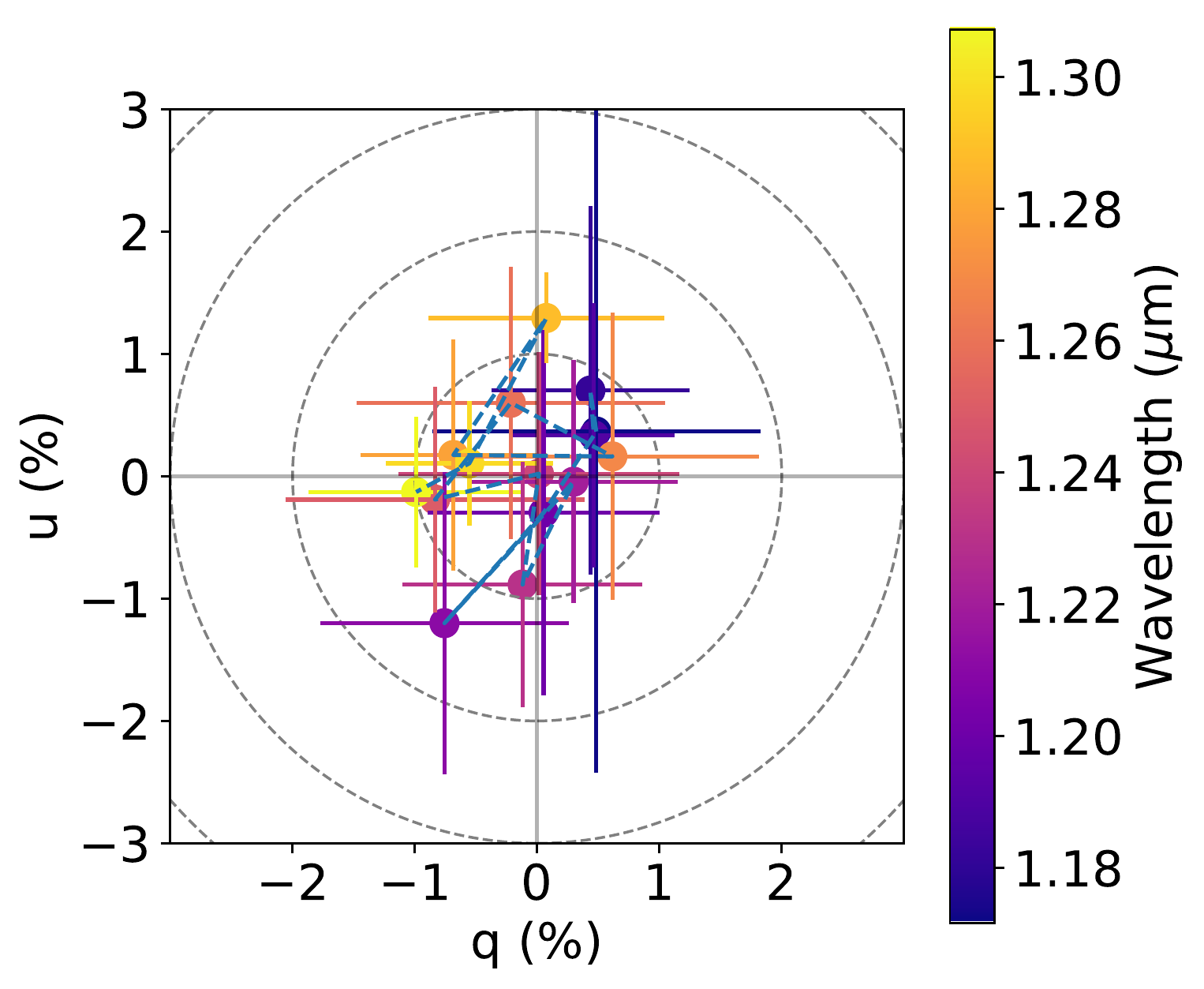}
    \caption{J-band spectropolarimetry of SN 2020qmp conducted using WIRC+Pol on October 29, {2020}. The panel shows the q-u plane, color coded by wavelength, and isolines of constant polarization.}
    \label{fig:polarization}
\end{figure}

Following the initial announcement of the discovery \citep{ATEL}, the transient was followed up by \textit{Swift} using the Ultra-Violet Optical Telescope \citep[UVOT;][]{Roming2005a} and the X-Ray Telescope \citep[XRT;][]{Burrows2005a}.
\newcommand{\swift}{\textit{Swift}} \swift\ observed the field with UVOT  between July 31 and August 8, 2020 (PI: Paraskeva). The brightness in the UVOT filters was measured with UVOT-specific tools in the High Energy Astrophysics Software package (HEASOFT)\footnote{\url{https://heasarc.gsfc.nasa.gov/docs/software/heasoft/}} version 6.26.1. Source counts were extracted from the images using a $3''$ radius aperture. The background was estimated using a circular region with a radius of $29''$ close to the SN position. The count rates were obtained from the images using the \swift\ tool uvotsource. They were converted to magnitudes using the UVOT photometric zero points \citep{Breeveld2011a}. {We investigated the extent to which contamination from the host galaxy affected the UV bands to see if corrections were necessary. On December 19, 2020, Swift observed the SN position in the UVM2 filter for 4.8 ks. We measured the host contribution at the SN position using the same apertures for the source and background region as for the SN. We then arithmetically subtracted the host flux from the SN LC. Before the host correction, the brightness in UVM2 varied between 15.2 and 15.7 mag. The brightness of the host galaxy at the SN position is UVM2 = 19.64 mag. The host is 4 mag fainter than the SN, meaning the host contributed less than 3\% to the emission at early times and is therefore negligible. Based on that, the host contribution is expected to be insignificant in W1 and W2 as well. In the optical filters, it is more difficult to estimate the host contribution without having templates in hand. It is likely small, but we cannot provide a quantitative estimate. Therefore, due to the lack of host templates, the SN flux includes the contribution from the host galaxy, and {the effects are minimal.}} All magnitudes were transformed into the AB system using \citet{Breeveld2011a}. 

\textit{Swift}/XRT observed the SN in the energy range from 0.3 to 10~keV. We analyzed all data with the online tools of the UK \swift\ team\footnote{\url{https://www.swift.ac.uk/user_objects/}} that use the methods described in \citet{Evans2007a} and \citet{Evans2009a} and HEASOFT. Combining the four epochs taken in July and August 2020 amounts to a total XRT exposure time of $3982$~s and provides a marginal detection of $0.0014^{+0.0009}_{-0.0007}$~count~s$^{-1}$ between 0.3 and 10 keV. If we assume a power-law spectrum with a photon index of $\Gamma = 2$ and a Galactic hydrogen column density of $1.95\times10^{20}$~cm$^{-2}$ \citep{HI4PI2016a}, this corresponds to an unabsorbed 0.3--10.0 keV flux of $5.1^{+3.3}_{-2.6}\times10^{-14}~{\rm erg\,cm}^{-2}\,{\rm s}^{-1}$. At the luminosity distance of SN 2020qmp, this corresponds to a luminosity of L$_{\rm{X}}=2\pm1\times10^{39}$~erg\,s$^{-1}$ (0.3--10~keV) on August 2, 2020 . A final 4.8 ks observation was obtained on December 19, 2020. {The source was not detected in X-rays}. The 3$\sigma$ count-rate limit is 0.002 ct s$^{-1}$. Using the same model as for the early-time observations, the luminosity is $<2.3\, \times 10^{39}\,{\rm erg\,s}^{-1}$ between 0.3 and 10 keV. 

The LC of the SN over a range of wavelengths is shown in Fig. \ref{LightcurveSpectra}. {We also calculated the bolometric LC of the SN {where a blackbody model provides a reasonable approximation to the photospheric spectrum of the SN. We performed a blackbody fit with all available filters for each photometric epoch and then integrated the blackbody to derive a luminosity.} For the early-time LC during the first 20 days after explosion, we used photometry within one-day windows to calculate our blackbody fits as the early-time LC is highly variable. After 20 days, we used photometry within four-day windows and then performed blackbody fits for every day that there was at least three different wavelength bands available.} The bolometric LC is presented in Fig. \ref{Bollc} and photometric measurements in Table \ref{MagnitudeTable}. {We note that though we display errors originating from individual photometry measurements in the LC, the overall error is dominated by the distance uncertainty to the source, which is 0.54 mag. We see that the temperature of the blackbody fits evolve from around 10,800 K at the beginning of the plateau to 5200 K at the end of the plateau, and we show this evolution over time in Fig. \ref{Bollc}}.

\begin{table}
\begin{tabular}{||c c c c||}
\hline
{Julian Date} & {Instrument} & {Filter} & {AB Magnitude} \\
\hline\hline
2459051.674& PGIR & J & > 15.26 \\ [0.5ex]
2459052.697 & ZTF & i & > 19.50 \\ [0.5ex]
2459056.673 & ZTF & r  & $14.6 \pm 0.03$ \\ [0.5ex]
2459056.678 & ZTF & r &  $14.56 \pm 0.03$ \\ [0.5ex]
2459056.683 & ZTF & r &  $14.59 \pm 0.03$ \\ [0.5ex]
2459060.662 & ZTF & r &  $14.29 \pm 0.03$ \\ [0.5ex]
2459060.664 & PGIR & J &  $14.74 \pm 0.02$ \\ [0.5ex]
\hline
\end{tabular}
\caption{{Summary of the photometric measurements obtained of SN 2020qmp, with apparent AB magnitudes reported. A full version of the table is available online.}}
\label{MagnitudeTable}
\end{table}

\subsection{Spectroscopy and classification}
\label{Spectroscopy}
\indent We initiated a rapid spectroscopic follow-up of the transient after the initial detection with the SED Machine (SEDM) spectrograph; \citep{Blagorodnova_2018, Rigault2019} on the Palomar 60-inch telescope (on July 31, 2020, 9 days post-explosion), the Gemini Multi-object Spectrograph on the {Gemini-North} telescope \citep{Gemini} (on July 31, 2020, 9 days post-explosion), and the Double Beam SPectrograph (DBSP; \citealt{Oke_1982}) on the Palomar 200-inch telescope (on August 12, 2020, 21 days post-explosion). {We show the spectral evolution in Fig. \ref{Spectra}}.
The presence of Balmer lines (H$\alpha$ and H$\beta$, labeled in Fig. \ref{Spectra}) points toward the classification of a Type II SN \citep{Filippenko1997, Gal-Yam2017}. Our spectra also show evidence for P-Cygni {profiles} from He I and Ca II. The relatively flat LC is characteristic of the plateau of constant brightness found in Type IIP SNe, typically expected to last around 100 days \citep{Branch2017}. With all this taken together, we classify SN 2020qmp as a Type IIP SN. Using the minimum of the strong P-Cygni profile of the H$\alpha$ line, we see that the photospheric velocities decrease over time. We {measured} a photospheric velocity of 9400 km $\mathrm{s}^{-1}$ in the first SEDM spectrum, 8800 km $\mathrm{s}^{-1}$ {in the Gemini spectrum}, and 7900 km $\mathrm{s}^{-1}$ {in the first P200 spectrum}.

We also obtained five additional optical spectra, one from SEDM (on August 28, 2020, 98 days post-explosion), one from the Low Resolution Imaging Spectograph (LRIS) on the Keck-I Telescope (\citealt{LRIS} on November 20, 2020, 121 days post-explosion), and three from DBSP (on January 8, 2021, 170 days post-explosion; on February 20, 2021, 212 days post-explosion; and on April 16, 2021, 267 days post-explosion). The last four spectra show the transition of the SN into the radioactive decay nebular phase, with characteristic nebular spectra features such as the [O I] doublet ($\lambda \lambda$ 6300, 6364 $\AA$), which strengthens with time as the SN progresses into the nebular phase (see {Sect. } \ref{Nebularsection}). 

Four spectra in the NIR were also obtained, one with the Near-Infrared Echellette Spectrometer {(NIRES) on the Keck Telescope} \citep[][]{NIRES} (on October 31, 2020, 101 days post-explosion), two with the Triple Spectrograph (TSPEC) on the Palomar 200-inch Telescope \citep[][]{TSPEC} (on December 12, 2020, 153 days post-explosion, and on February 4, 2021, 197 days post-explosion), and one with the NASA Infrared Telescope Facility SpeX (IRTF + SpeX) instrument \citep{IRTF} (on May 16, 2021, 298 days post-explosion) as part of program 2020A111 (PI: K. De). All obtained spectra are shown in Fig. \ref{Spectra}. 

{The spectra were all reduced using standard techniques, including wavelength calibration with
arc-lamp spectra and flux calibration using spectrophotometric standard stars. SEDM was reduced through the fully automated Python-based reduction pipeline pysedm6 \citep{Rigault2019}, Gemini through the Gemini Image Reduction and Analysis Facility (IRAF) package following procedures detailed in the Gemini Multi-Object Spectrographs Data Reduction Cookbook, DBSP through the Python-based IRAF reduction pipeline \citep{Bellm2016}, LRIS through the IDL-based reduction pipeline LPipe \citep{Perley2019}, and NIRES, TSPEC, and IRTF through the IDL-based data reduction package Spextool9 \citep{Cushing2004}. Corrections for NIR telluric absorption features from the Earth's atmosphere were done using the method developed by \citet{Vacca2003}.}

\subsection{Near-infrared spectropolarimetry}
The proximity {and brightness} of SN\,2020qmp allowed for spectropolarimetric observations in the infrared. 
Such observations can constrain the geometry of the ionized electron-scattering region in the SN.
Near-infrared spectropolarimetry has the added benefit of less contamination from dust polarization along the line of sight, both in the host galaxy and in the Milky Way \citep{Nagao2018}. 
We observed the SN on October 29, 2020, 99 days post-explosion, while the SN was still in the plateau phase, with the apparent magnitude of $J = 13.2 \pm 0.13$ mag.
The observation was obtained using the infrared spectropolarimeter WIRC+Pol on the 200-inch telescope at Palomar Observatory \citep{Tinyanont2019a}. 
The SN was observed inside its $3''$ wide slit in an ABAB dithering pattern for a total of 64 min of exposure time. 
The observations were performed at high airmass (average of 1.8), resulting in low flux due to the large atmospheric extinction.
WIRC+Pol {exhibits} $<$0.03\% instrumental polarization, and observations of unpolarized standard stars were not necessary \citep{Tinyanont2019a}.
The data were reduced using the WIRC+Pol data reduction pipeline.\footnote{\url{https://github.com/WIRC-Pol/wirc_drp}}

Figure~\ref{fig:polarization} shows the normalized Stokes parameters (values describing the state of polarization present in the electromagnetic radiation being studied) $q$ and $u$ plotted against each other, color coded by wavelength.
On this plot, the distance from origin is the degree of polarization, $p$, while the angle with respect to the $x$ axis is twice the angle of polarization, $\theta$.
We did not detect polarization from SN\,2020qmp, as the broadband degree of polarization was $0.14 \pm 0.26 \%$, making the SN unpolarized to within $0.78\%$ at the 3$\sigma$ level.
The typical error bar per spectral channel is 1\% in both $q$ and $u$, and the broadband upper limits are 0.25\% and 0.27\% in $q$ and $u$, respectively (all 1$\sigma$). 
The non-detection of polarization of a Type {IIP} SN during the plateau phase {is consistent with most Type {IIP} SNe} because the outer ejecta, visible during this phase, are generally symmetric (see the review by \citealp{Wang2008}).

\subsection{Observations by the VLA}
\label{radiosection}

\begin{figure}
    \centering
    \includegraphics[width = 0.8\linewidth]{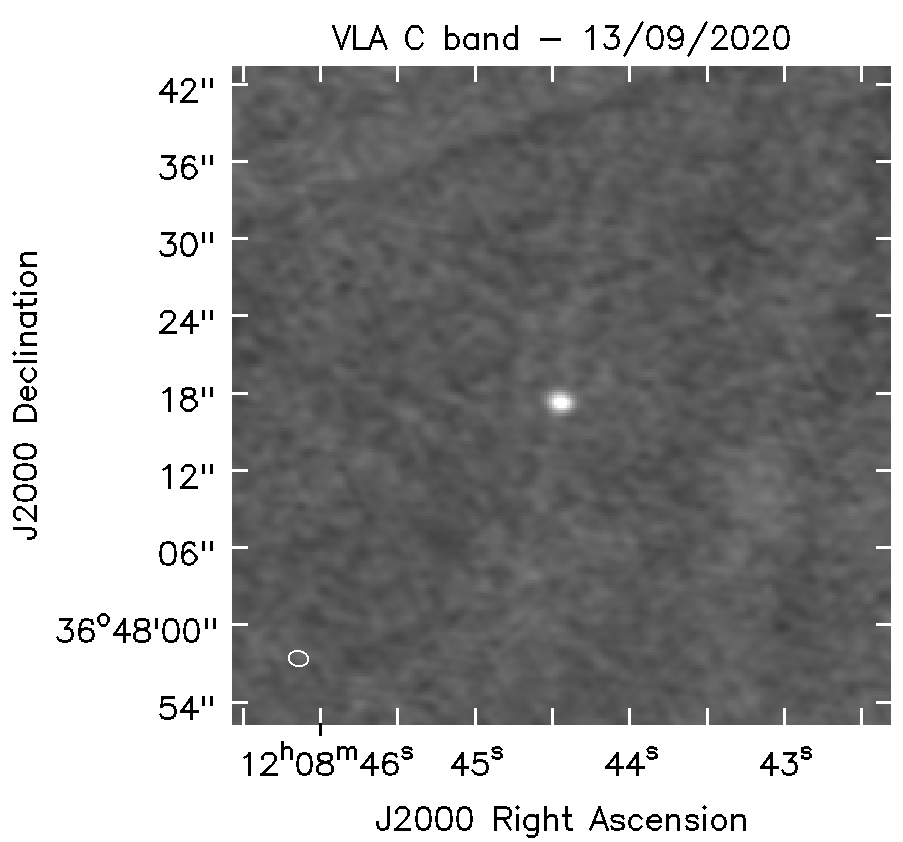}
    \caption{Image of SN 2020qmp in the C band taken by the VLA on September 13, 2020.}
    \label{radiofigure}
\end{figure}

\begin{table}
\begin{tabular}{||c c c c||}
\hline
{$\Delta t$} \textrm{(Days)} & {$\nu$} \textrm{(GHz)} & {$F_{\nu}$} \textrm({mJy)} & {Configuration} \\
\hline\hline
$51$ & $5$ & $0.32 \pm 0.04$ &  B \\ [0.5ex]
$51$ & $7$ & $0.19 \pm 0.02$ &  B \\ [0.5ex]
$51$ & $22$ & $0.08 \pm 0.02$ &  B \\ [0.5ex]
$57$ & $2.5$ & $0.26 \pm 0.05$ &  B \\ [0.5ex]
$57$ & $3.5$ & $0.26 \pm 0.04$ &  B \\ [0.5ex]
$57$ & $5$ & $0.25 \pm 0.03$ &  B \\ [0.5ex]
$57$ & $7$ & $0.19 \pm 0.03$ &  B \\ [0.5ex]
$57$ & $10$ & $0.12 \pm 0.02$ & B \\ [0.5ex]
$57$ & $15$ & $0.08 \pm 0.02$ & B \\ [0.5ex]
$104$ & $2.31$ & $0.59 \pm 0.09$ & $\rm{B=>A}$ \\ [0.5ex]
$104$ & $2.94$ & $0.39 \pm 0.06$ &  $\rm{B=>A}$ \\ [0.5ex]
$104$ & $3.63$ & $0.33 \pm 0.05$ &  $\rm{B=>A}$ \\ [0.5ex]
$104$ & $5$ & $0.22 \pm 0.03$ &  $\rm{B=>A}$ \\ [0.5ex]
$104$ & $7$ & $0.14 \pm 0.03$ &  $\rm{B=>A}$ \\ [0.5ex]
$104$ & $10$ & $0.09 \pm 0.02$ &  $\rm{B=>A}$ \\ [0.5ex]
$104$ & $15$ & $0.06 \pm 0.01$ &  $\rm{B=>A}$ \\ [0.5ex]
$136$ & $2.31$ & $0.33 \pm 0.05$ & $\rm{BnA=>A}$ \\ [0.5ex]
$136$ & $2.94$ & $0.31 \pm 0.04$ &  $\rm{BnA=>A}$ \\ [0.5ex]
$136$ & $3.63$ & $0.24 \pm 0.05$ &  $\rm{BnA=>A}$ \\ [0.5ex]
$136$ & $6$ & $0.13 \pm 0.02$ &  $\rm{BnA=>A}$ \\ [0.5ex]
$136$ & $10$ & $0.07 \pm 0.01$ &  $\rm{BnA=>A}$ \\ [0.5ex]
$136$ & $15$ & $0.04 \pm 0.01$ &  $\rm{BnA=>A}$ \\ [0.5ex]
\hline
\end{tabular}
\caption{Summary of the radio observations of SN 2020qmp. $\Delta t$ is the {time from the estimated explosion date}. $\nu$ is the observed frequency in GHz. {The fluxes are consistent with the large range of values found for Type IIP events \citep{Bietenholtz}.} }
\label{fluxtable}
\end{table}

\indent The VLA observed the field of SN 2020qmp (under our Director's Discretionary Time program VLA/20B-398; PI Horesh) and detected radio emission consistent with the SN position in four epochs. The first observation, on September 13, 2020, showed a point source in both the C band (6 GHz) and the K band (22 GHz) at a flux level of $0.25$ and $0.08$ mJy, respectively. The detection image in the C band is shown in Fig. \ref{radiofigure}. We continued monitoring the SN with the VLA using the S, C, X, and Ku bands (3, 6, 10, and 15 GHz) for three additional epochs up to 136 days post-explosion.

\indent We calibrated our observations with the automated VLA calibration pipeline available in the Common Astronomy Software Applications (CASA) package \citep{McMullin2007}. 3C286 was used as the primary flux calibrator, while J1146+3958 was used as the gain calibrator. When imaging the field of SN 2020qmp with the CASA task CLEAN, we divided {the C band into two sub-bands and the S band into two or three sub-bands, when the signal-to-noise ratio was high enough.} We used the CASA task IMFIT to fit the source in the phase center and to extract the peak flux density. We estimated its error as the square root of the quadratic sum of the error produced by the CASA task IMFIT, the image rms produced by the CASA task IMSTAT, and the 10\% calibration error. We report the flux density measurement in Sect.  \ref{progenitor} and in Table \ref{fluxtable}.

\section{Nebular spectrum analysis}
\label{Nebularsection}
After the photosphere recedes into the ejecta after the hydrogen recombination plateau phase ends, the ejecta become optically thin in the continuum and the inner regions become visible, providing insights into the nucleosynthesis {in the explosion}. During this phase, the luminosity becomes directly proportional to how much $^{56} \rm{Ni}$ was created during the explosion. A spectrum taken in this phase allows us to infer the nucleosynthetic yields of the explosion, which allows for the measurement of the ZAMS progenitor mass through the comparison of line strengths with existing models as nucleosynthesis is strongly dependent on the mass of the progenitor. This phase of the LC is called the nebular phase, where the powering of the LC becomes dominated by the radioactive decay of $^{56}\mathrm{Co}$. In particular, comparison of the intensities of the [O I] doublet has been shown to provide a good indication of the ZAMS mass \citep{Uomoto1986, Jerkstrand2014}, which we use in our analysis.  

\citet{Jerkstrand2014} developed the models that we use in our analysis. They started with evolved ejecta exploded using the hydrodynamic code KEPLER \citep{Woosley2007} and created the spectra using the radiative transfer code CMFGEN \citep{Jerkstrand2011}. The models are computed at different time epochs after the explosion, {for ZAMS masses of 9, 12, 15, 19, and 25 $\mathrm{M_\odot}$, provided by \citet{Jerkstrand2014} and \citet{Jerkstrand2018}}. Using these models, along with our observed spectra, we then estimated the ZAMS mass of the progenitor star {by} first calculating the $^{56}\rm{Ni}$ mass and then comparing the [O I] doublet line luminosity normalized relative to the $^{56} \mathrm{Co}$ decay power. 
\subsection{$^{56} {Ni}$ mass calculation}
 It is well known that the nebular phase of Type II SNe is powered through the nuclear radioactive decay of $^{56} \mathrm{Ni}$ to $^{56} \mathrm{Co}$, and then to $^{56} \mathrm{Fe}$. During this process, $\gamma$ rays and positrons are released; however, at this point the ejecta are still not transparent to $\gamma$ rays, and as a result the bolometric luminosity during the nebular phase can be used to determine the $^{56} \mathrm{Ni}$ mass through the relation \citep{Spiro2014}
 \begin{equation}
    M_{SN}(Ni) = 0.075 \times  L_{SN}/L_{87A} \, \mathrm{M_\odot} \, ,
\label{eq1}
\end{equation}
where $L_{SN}$ is the bolometric luminosity of the SN in question, and $L_{87A}$ is the bolometric luminosity of SN 1987A. 

\begin{figure}
    \centering
    \includegraphics[width = 1.1\linewidth]{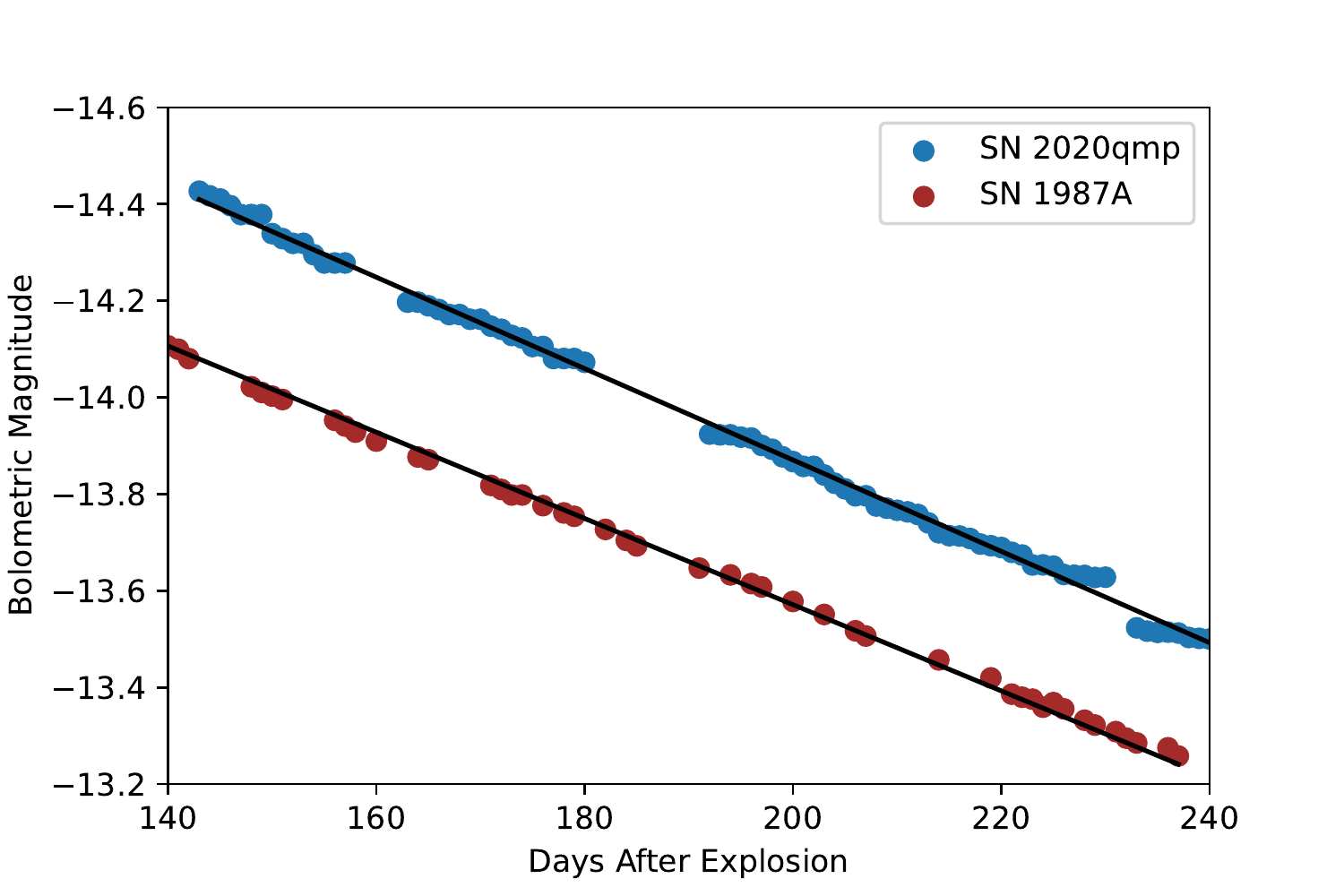}
    \caption{Zoomed-in view of the radioactive decay tail of the bolometric LCs of SN 2020qmp and SN 1987A. {Best-fit lines fit to the LCs used to derive the $^{56}\mathrm{Ni}$ masses are also shown.} }
    \label{bolometric}
\end{figure}

{To make this comparison, we computed pseudo-bolometric LCs of SN 2020qmp and SN 1987A over the $g$, $r$, and $i$ filters during the radioactive decay phase. The LC of SN 1987A\footnote{LC compiled from the Open Supernova Catalog: https://sne.space \citep{OpenSupernovaCatalog}} was measured over the B, V, R, and I filters, and we used the transformations presented by  \citet{Chonis2008} to convert the magnitudes into the $g$, $r$, and  $i$ bands to allow for a direct comparison with SN 2020qmp. The pseudo-bolometric LCs were calculated by using the trapezoidal numerical integration method over the entire spectral region that encompasses the three bands. We show the pseudo-bolometric LCs in Fig. \ref{bolometric}, in absolute AB magnitude space, along with characteristic best-fit lines fit to the decay phases. We obtain slopes corresponding to a decay rate of about 0.01 AB mag per day for both SNe. This is a decay rate consistent with complete gamma-ray trapping of the ejecta \citep{Arnett1980}, which confirms that we can use Eq. \ref{eq1} to reliably estimate the $^{56}Ni$ mass. We then converted the difference of the y intercepts of the two best-fit lines into a luminosity ratio and used Eq. \ref{eq1} to obtain $M_{SN}(\mathrm{Ni}) =  0.11^{+0.07} _{-0.04} \, \mathrm{M_\odot}$.  Though the epoch we use to compare the luminosities does not matter, we compared the y intercepts of our best-fit lines in order to average out any random noise present in the observations. We note that the uncertainty obtained is due primarily to the uncertainty on the distance to the SN.}

\subsection{Normalized [O I] line luminosity and ZAMS mass}
\begin{figure}
    \centering
    \includegraphics[width = 1.1\linewidth]{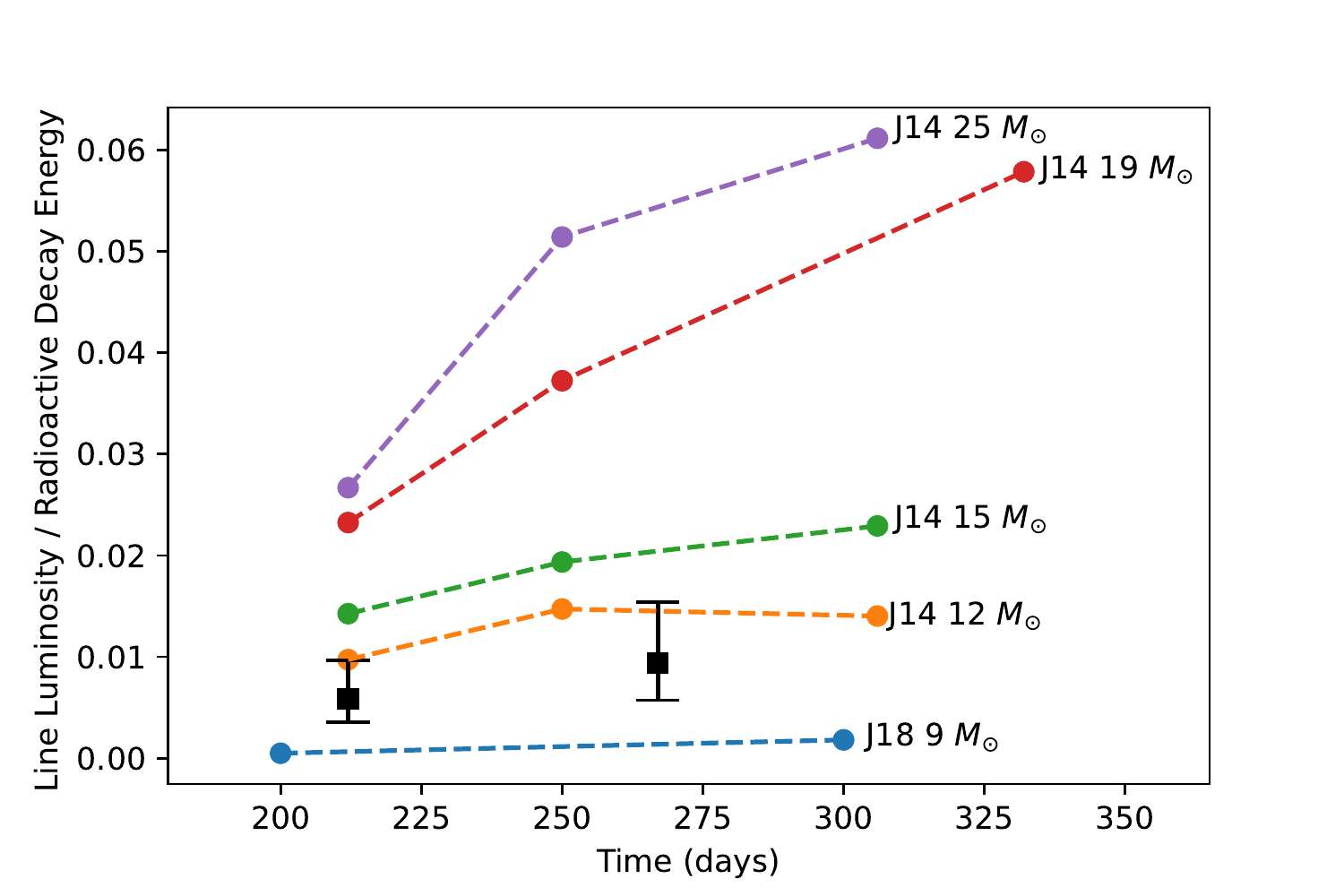}
    \caption{Normalized line luminosities of the [O I] doublet ($\lambda \, 6300 \,, 6364 \,  \AA \,$) at different time epochs for the observed spectra of SN 2020qmp as well as the models from \citet{Jerkstrand2014} and \citet{Jerkstrand2018}. The observed luminosities from SN 2020qmp are plotted as squares, with their error bars in black; the models are each plotted with circles, with different colors corresponding to the different models.}
    \label{Normalizedlum}
\end{figure}

After obtaining the $^{56} \mathrm{Ni}$ mass, we then calculated the normalized [O I] doublet line luminosity, relative to the $^{56} \mathrm{Co}$ decay power, which is the main characteristic used to compare model spectra to the observed spectrum. The normalized luminosity is given by \citet{Jerkstrand2015} as
\begin{equation}
    L_{\mathrm{norm}}(t) = \frac{L_{\mathrm{line}}}{1.06 \times 10^{42} \frac{M_\mathrm{Ni}}{0.075 \mathrm{M_\odot}}(e^{-t / 111.4 d} - e^{-t/8.8 d}) \, \mathrm{erg \, s^{-1}}} .
\label{lumeq}
\end{equation}
Using the $^{56} \mathrm{Ni}$ mass obtained from the photometry, as well as the line luminosity for the [O I] doublet ({by subtracting the continuum from the spectrum and integrating the line flux by fitting a double-peaked Gaussian function), }we obtained normalized line luminosities at 211 days and 266 days after the explosion through Eq. \ref{lumeq} ($0.006 ^{+0.004}_{-0.002}$ and $0.01 ^{+0.006}_{-0.004}$). We also note that before calculating these luminosities, we scaled the spectra to match the photometry obtained by ZTF at the same epoch in order to get an absolute flux calibration. Then, going through the same process of fitting a double-peaked Gaussian for each of our model spectra, and assuming the same $^{56} \rm{Ni}$ mass, we also obtained normalized line luminosities for each model spectra and compare the results in Fig. \ref{Normalizedlum}. We see that the normalized luminosities obtained for the observed spectra {are between the 9 $\mathrm{M_\odot}$ and 12 $\mathrm{M_\odot}$ models. This allows us to infer that the progenitor star is between 9 and 12 $\mathrm{M_\odot}$}.

\section{Hydrodynamical LC modeling}
\label{LCModeling}

It is also possible to constrain the ZAMS progenitor mass and initial explosion energy of the SN through hydrodynamical LC modeling \citep{Utrobin2015, Utrobin2017, Morozova2017, Morozova2018, Goldberg2019, Martinez2019}. In order to do so, we used the open-source SN Explosion Code \citep[SNEC;][]{Morozova2015}. This code assumes local thermodynamic equilibrium and diffusive radiative transport, allowing it to model LCs well up to the radioactive decay phase, where these assumptions break down. Therefore, we only compared the model LCs generated through SNEC to the observed LCs up to 125 days after the explosion, when the plateau phase has noticeably transitioned to the {optically thin nebular phase}. 

\begin{figure}
    \centering
     \includegraphics[width = 1\linewidth]{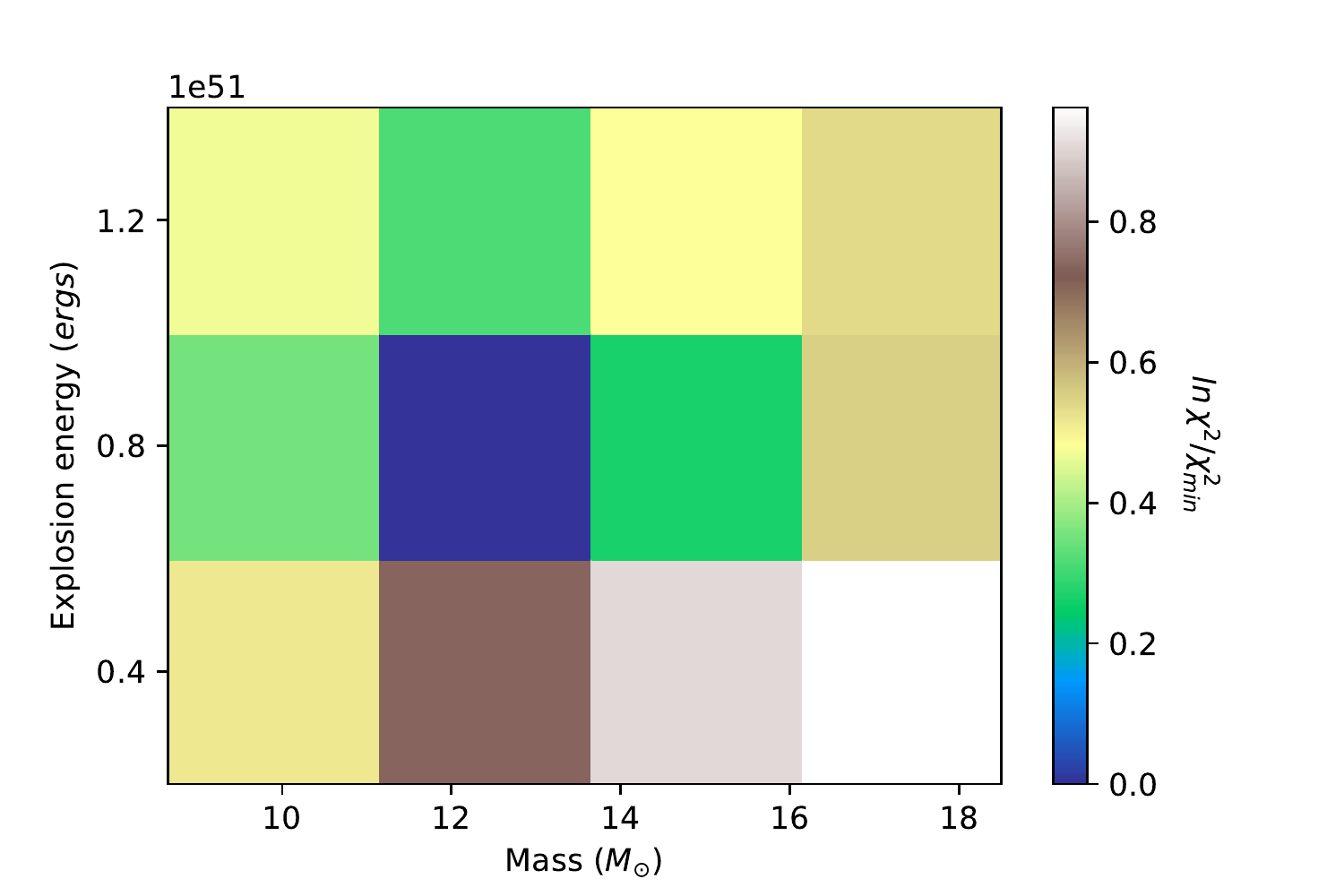}
     \includegraphics[width = 1\linewidth]{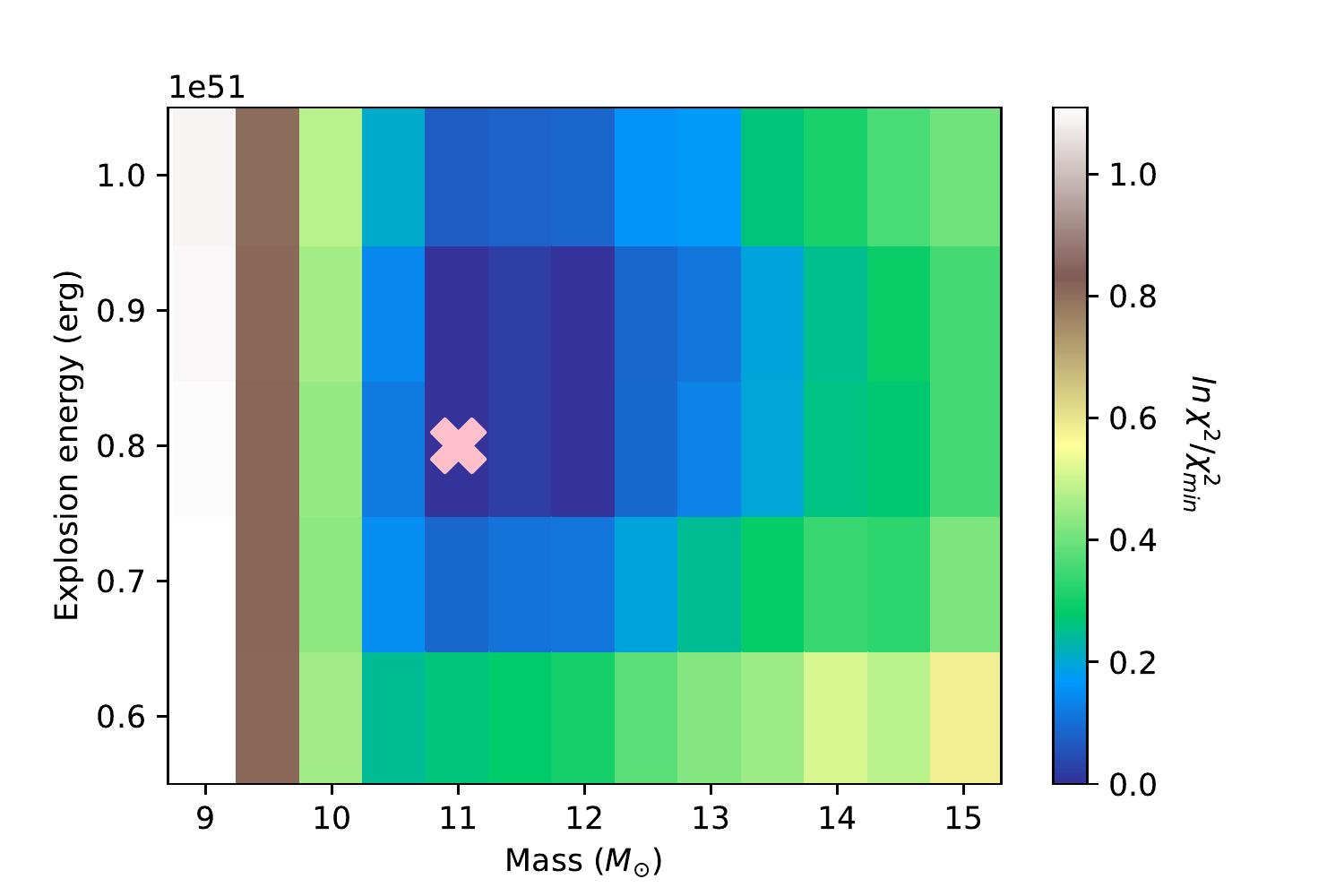}
    \caption{Plots showing the $\chi^2$ values for different subsets of hydrodynamical LC models compared to the observed LCs. \textit{Top panel:} $\chi^2$ values for the different LC models compared to observations, over a range of 10 to 17.5 $\mathrm{M_\odot}$ ($\Delta M = 2.5 \, \mathrm{M_\odot}$) and an explosion energy of 0.4 to 1.2 $\times \, 10^{51}$ erg ($\Delta E = 0.4 \times 10^{51}$). \textit{Bottom panel:}  $\chi^2$ values for the different LC models compared to observations, over a range of {9 to 15} $\mathrm{M_\odot}$ ($\Delta M = 0.5 \, \mathrm{M_\odot}$) and an explosion energy of 0.6 to 1.0 $\times \, 10^{51}$ ergs ($\Delta E = 0.1 \times 10^{51}$), with an "X" demarcating the best-fit model. }
    \label{bestmodel}
\end{figure}

\begin{figure}
    \centering
    \includegraphics[width = 1\linewidth]{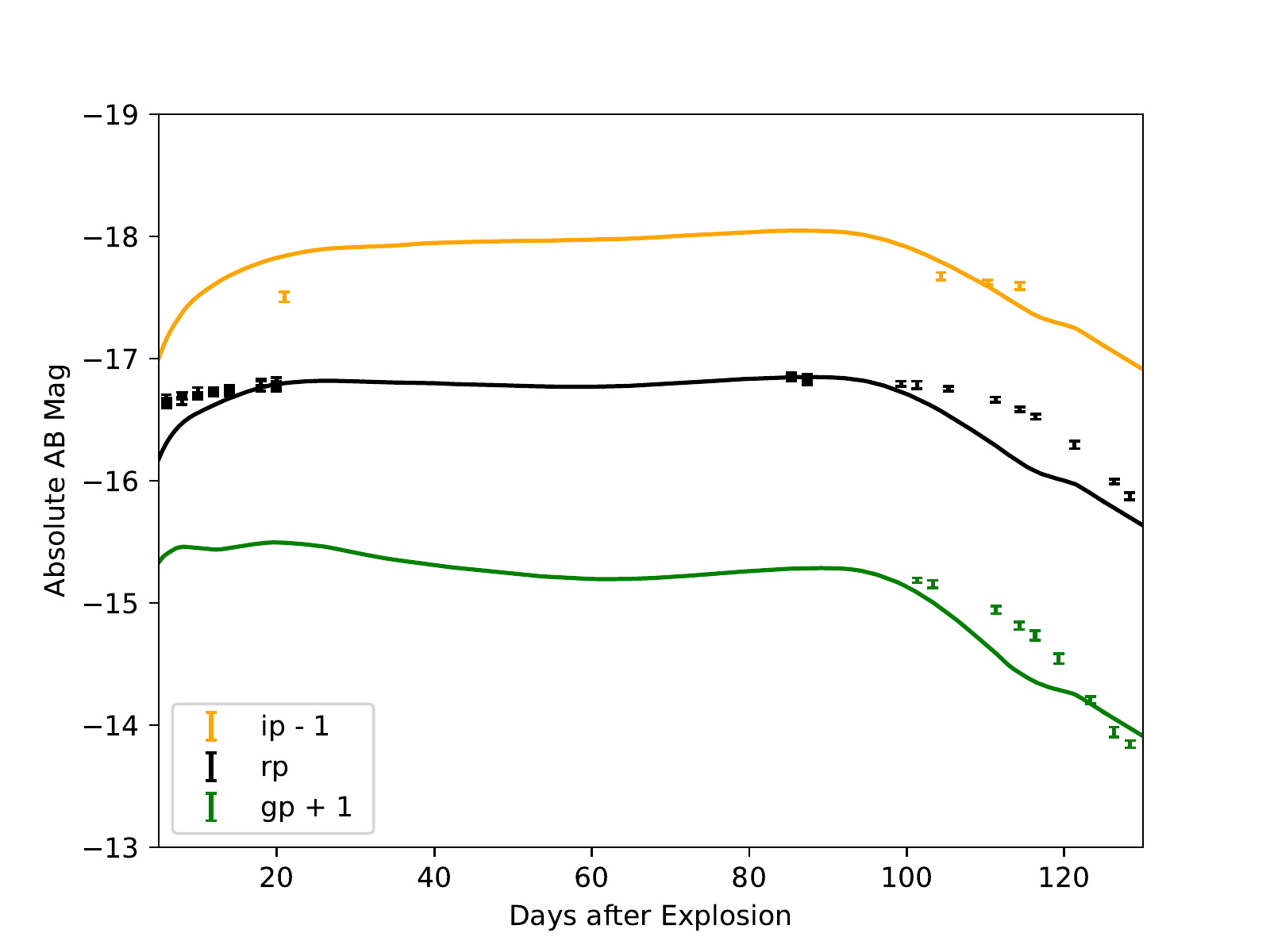}
    \caption{Best-fit LC models in the $i$, $r$, and $g$ bands, representing a progenitor star of 11.0 $\mathrm{M_\odot}$ and an initial explosion energy of 0.8 $\times$ $10^{51}$ ergs. The models were fit to the observed LC up to 125 days after the explosion.}
    \label{Bestfitmodel}
\end{figure}
In our analysis, we followed an approach similar to that in \citet{Morozova2017} and {began by} generating optical LCs corresponding to a range of ZAMS {progenitor} masses between 10.0 $\mathrm{M_\odot}$
and 17.5 $\mathrm{M_\odot}$ ($\Delta M = 2.5 \, \mathrm{M_\odot}$) and a range of explosion energies between {0.4} and 1.2 $\times \, 10^{51}$ erg ($\Delta E = 0.4 \times \, 10^{51} \, \mathrm{erg}$) with stellar structure models obtained from \citet{Sukhbold2016}. We fixed the $^{56} \mathrm{Ni}$ mass of the models to $M_\mathrm{Ni} = 0.11 \, \mathrm{M_\odot}$, which we obtained using photometry on the radioactive decay tail (see Sect. \ref{Nebularsection}). {We then ran through a coarse grid} of models and performed a $\chi^2$ analysis between the models and the observed LCs in the $g$, $r$, and $i$ bands. {This $\chi^2$ analysis is done by comparing every observed photometry point from the ZTF LC to the model point equivalent in time to the observed point, corresponding to the same band filter. The models generate photometry points in units of absolute AB magnitudes, so we converted the model points to fluxes and performed the $\chi^2$ analysis in flux space.} We {found} that the best fitting models are around 12.5  $\mathrm{M_\odot}$ and around an explosion energy of $0.8 \times 10^{51}$ erg. We then {ran} the SNEC code with a finer parameter space, between 9 and 15 $\mathrm{M_\odot}$ ($\Delta M = 0.5 \, \mathrm{M_\odot}$) and between 0.6 and 1.0 $\times 10^{51}$ erg ($\Delta E = 0.1 \times 10^{51} \,  \mathrm{erg}$), and repeated the $\chi^2$ analysis to obtain the final best fitting progenitor mass as well as explosion energy. The results of the analysis are shown in Fig. \ref{bestmodel}, and we obtain a {best-fit ZAMS progenitor mass of 11.0 $\mathrm{M_\odot}$ and an initial explosion energy of $0.8 \times 10^{51}$ erg.} The best-fit model LCs along with the observed LCs are presented in Fig. \ref{Bestfitmodel}. This best-fit mass is within the range of the ZAMS progenitor mass obtained using the independent method in Sect.  \ref{Nebularsection} with the nebular spectra, {and it is better constrained}.

\section{Modeling the radio data combined with optical and X-ray}
\label{progenitor}

\begin{figure}
\centering
\includegraphics[width = \linewidth]{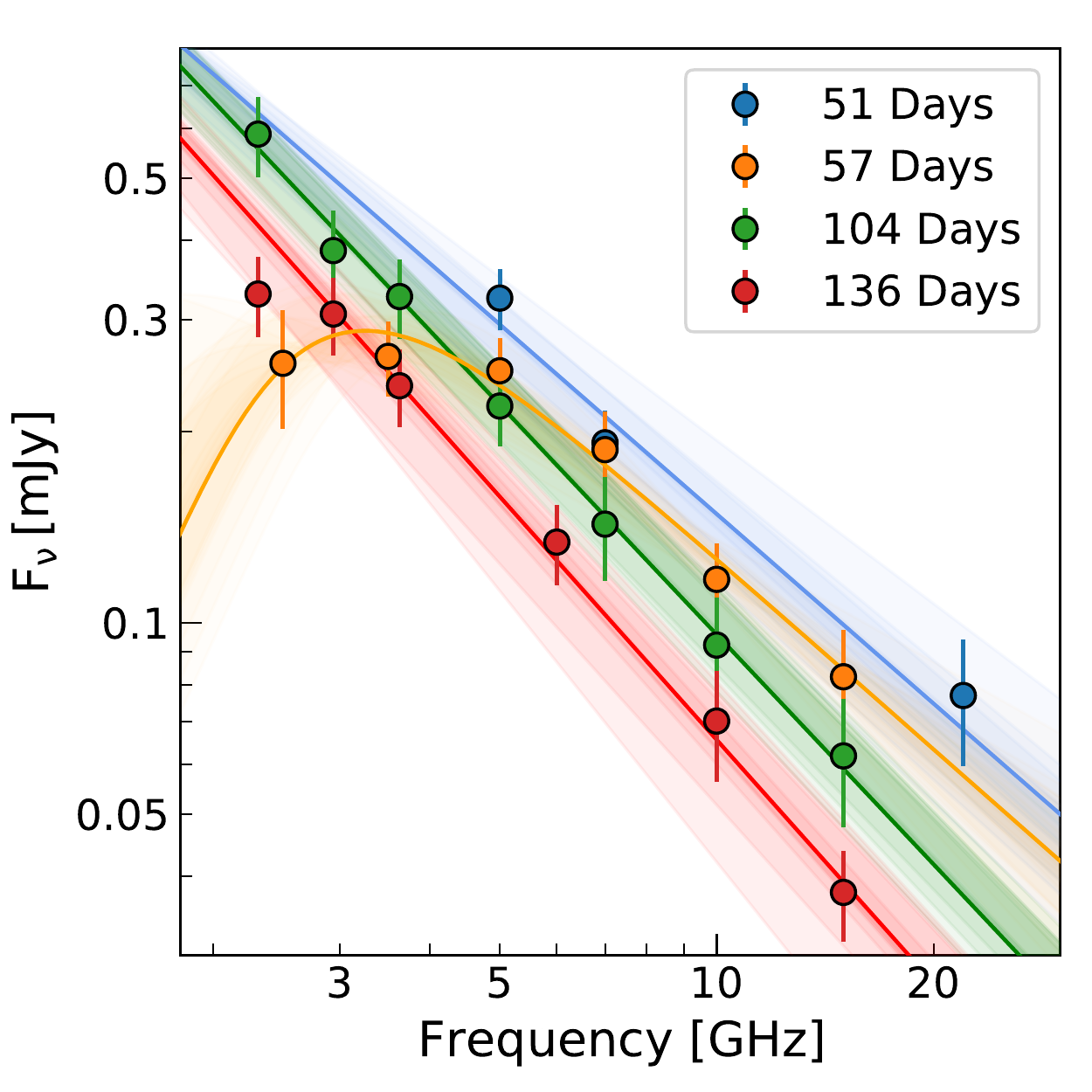}
\caption{VLA radio spectral energy distributions of SN 2020qmp at four different epochs. The lines are fitted models as discussed in Sect.  \ref{subsec: radio_modeling}, {with shaded confidence regions of $1 \sigma$ drawn from the posterior distributions.} A two-power-law model, as presented in Eq. 4 in \cite{Chevalier1998}, was fitted to the spectrum 57 days after the explosion. An optically thin power-law model was fitted to the spectra 51, 104, and 136 days after the explosion.}
\label{radio_spectra}
\end{figure}

\indent The radio spectra obtained with the VLA (as described in Sect.  \ref{radiosection}) are presented in Fig. \ref{radio_spectra}. The SN exhibits an optically thin emission $51$ days after the explosion (at $\geq 5$\,GHz; no lower frequencies were observed at that epoch), while the 57-day spectrum shows a turnover at around 4 GHz into an optically thick spectrum. However, the turnover frequency is not well constrained due to scarce data in the optically thick regime. On day $104$, an optically thin spectrum is observed down to a frequency of $2.31$\,GHz, surprisingly at a significantly higher flux density at frequencies lower than the turnover frequency observed earlier. The last spectrum, $136$ days after the explosion, exhibits an optically thin emission with a possible turnover at the lowest observed frequency, {at around 2.3 GHz}. This turnover frequency is even less constrained than the one at $57$ days after the explosion. In the following section, we discuss the radio data in light of a SN-CSM interaction model and derive the shock physical parameters (radius and magnetic field strength), the inferred shock velocity, and the progenitor's mass-loss rate. We also discuss the possibility of a variable CSM density structure, as suggested by the temporal evolution of the radio spectrum. Then, we use the X-ray and optical data combined with the radio to estimate the shock micro-physical parameters and discuss their effects on our estimates of the shock properties.

\subsection{Modeling the radio spectra}
\label{subsec: radio_modeling}

When the SN ejecta interacts with the CSM, it drives a shockwave into the CSM. At the shock front, electrons are accelerated to relativistic velocities with a power-law energy density distribution of $N(E) \propto E^{-p}$, {where $p$ is the electron spectral index}. The magnetic field is also enhanced at the shock front. The relativistic electrons that gyrate in the presence of that magnetic field give rise to synchrotron emission, which is usually observed at radio frequencies \citep{Chevalier1982}. The intrinsic synchrotron emission might be absorbed by synchrotron self-absorption (\citealt{Chevalier1998}) and/or free-free absorption (\citealt{Weiler_2002}). The optically thin regime of the spectrum is expected to follow a power-law function ($F_{\nu} \propto \nu^{-\beta}$); when in {the} absence of cooling (inverse Compton cooling), we expect a constant power law, $\beta = (p-1)/2$. The full shape of the spectrum as a function of the radio-emitting shell radius and the magnetic field strength is shown in Eq. 1 in \cite{Chevalier1998}.

\cite{Chevalier2006} have shown that in the case of a synchrotron self-absorbed spectrum, the radius of the synchrotron-emitting shell and the magnetic field strength can be obtained when the radio spectral peak is observed. For a spectrum with an observed peak flux density $F_{\nu_a}$ at a frequency $\nu_a$, assuming a typical power-law index $p=3$ \citep{Chevalier1998}, the radius is given by
\begin{multline}
    R = 4.0 \times 10^{14} \alpha^{-1/19} \left(\frac{f}{0.5}\right)^{-1/19}\left(\frac{F_{\nu_a}}{\mathrm{mJy}}\right)^{9/19}
    \\
    \times \left(\frac{D}{\mathrm{Mpc}}\right)^{18/19}\left(\frac{\nu_a}{5\,  \mathrm{GHz}}\right)^{-1} \mathrm{cm}\, , \label{Eq: chev_radius}
\end{multline}
where $D$ is the distance to the SN, $f$ is the emission filling factor, and the equipartition parameter, $\alpha$, is the ratio between the fraction of energy deposited by the shock to the relativistic electrons ($\epsilon_e$) and the magnetic field ($\epsilon_B$) \citep{Chevalier2006}. The magnetic field strength, in this case, is given by
\begin{multline}
    B = 1.1 \, \alpha^{-4/19} \left(\frac{f}{0.5}\right)^{-4/19}\left(\frac{F_{\nu_a}}{\mathrm{mJy}}\right)^{-2/19}
    \\
    \times \left(\frac{D}{\mathrm{Mpc}}\right)^{-4/19}\left(\frac{\nu_a}{5\,  \mathrm{GHz}}\right) \, \mathrm{G}\, . \label{Eq: chev_magnetic}
\end{multline}

We first modeled the optically thin spectra at $51$, $104$, and $136$ days after the explosion as power-law functions of the form $F_{\nu} \sim \nu ^{-\beta}$. We did not use the data at the lowest frequency on day $136$ as it might feature a turnover (and thus a deviation from a simple power-law function). We {performed} a $\chi^2$ minimization fit. For the first spectrum ($51$\,days), the fit resulted in $\beta = 0.99 \pm 0.06$ with a minimum $\chi^2 = 1.58$ and one degree of freedom (dof). However, this is based only on three data points and therefore should be treated carefully. For the spectrum on day $104$, we find $\beta = 1.20 \pm 0.06$ with a minimum $\chi^2$ of $0.5$ (and five dof). The fit of the last epoch ($136$\,days) resulted in a power law of $\beta = 1.27 \pm 0.06$ with a minimum $\chi^2$ of $0.34$ (and three dof). The above fits, with their $1 \sigma$ confidence interval, are shown in Fig. \ref{radio_spectra}. These power laws correspond, in the non-cooling regime, to $p = 2.98 \pm 0.12$, $3.40 \pm 0.12$, and $3.54 \pm 0.12$, for the first, third, and fourth spectrum, respectively. However, the actual value of $p$ will differ if cooling effects are taking place; in other words, if its real value is $p=3,$ then the rather steep spectral slopes are due to cooling.


To derive the shock physical parameters, we fitted a parameterized model, similar to Eq. 4 in \cite{Chevalier1998}, to the spectrum observed at $57$ days after the explosion. The free parameters are the peak flux density, $F_{\nu_a}$, its frequency, $\nu_a$, and the spectral index of the optically thin regime, $\beta$. We used \texttt{emcee} \citep{Foreman_Mackey_2013} to preform a Markov chain Monte Carlo analysis to determine the posteriors of the parameters of the fitted model (and used flat priors). We find, for the spectrum taken $57$ days after explosion, a peak flux density of $F_{\nu_a} = 0.28 \pm 0.03$ mJy at $\nu_a = 2.9^{+0.5} _{-0.7}$ GHz and an optically thick power law of $\beta = 1.0 \pm 0.2$. Given these fitted parameters and assuming $p=3$, $f=0.5$, and equipartition $\left( \alpha = \epsilon_e / \epsilon_B = 1 \right)$, the radius of the emitting shell is $R = \left( 5.1 ^{+1.8} _{-1.0} \right) \times 10^{15}$ cm and the magnetic field strength is $B = 0.4 \pm 0.1$ G. Assuming a constant shock velocity -- $v_{sh} = R/t$, where $t$ is the time since explosion -- we derive $v_{sh} = 1.0^{+0.4} _{-0.2} \times  10^4$ $\rm{km \, s^{-1}}$ on day $57$ after the explosion.



We next assumed that the CSM, shocked by the SN ejecta, was deposited via mass loss from the progenitor star before the explosion. Thus, the radio emission modeling can be used to estimate a mass-loss rate, assuming a constant mass-loss rate via a constant stellar wind velocity. Under this assumption, the CSM density structure has the form of $\rho \sim \frac{\dot{M}}{v_w} r^{-2}$, where $\mathrm{\dot{M}}$ is the mass-loss rate and $v_w$ is the wind velocity. Assuming that the magnetic field energy density is a fraction, $\epsilon_B$, of the post-shock energy density, $\sim \mathrm{\rho v_{sh}^2}$, and a constant shock velocity, the mass-loss rate is given by
\begin{multline}
    \mathrm{\dot{M}} = 5.2 \times 10^{-8} \left(\frac{\epsilon_B}{0.1}\right)^{-1} \left(\frac{\mathrm{B}}{\mathrm{1 \, G}}\right)^{2} \left(\frac{\mathrm{t}}{10 \, \mathrm{Days}}\right)^{2}
    \\
    \times \left(\frac{\mathrm{v_w}}{10 \, \mathrm{km/s}}\right) \, \mathrm{M_{\odot} \, yr^{-1}}\, . \label{Eq: chev_mass_loss}
\end{multline}
Thus, assuming $\epsilon_B = 0.1$, the mass-loss rate derived from the fitted model at $57$ days after explosion is $\mathrm{\dot{M}} = \left( 2.9^{+1.0} _{-1.2} \right) \times 10^{-7} \, \rm M_{\odot}\,  yr^{-1}$ for an assumed wind velocity of $10$ km/s.

\subsection{A variable CSM density structure}
\label{subsec: variable_CSM}

\begin{figure}
    \centering
    \includegraphics[width = \linewidth]{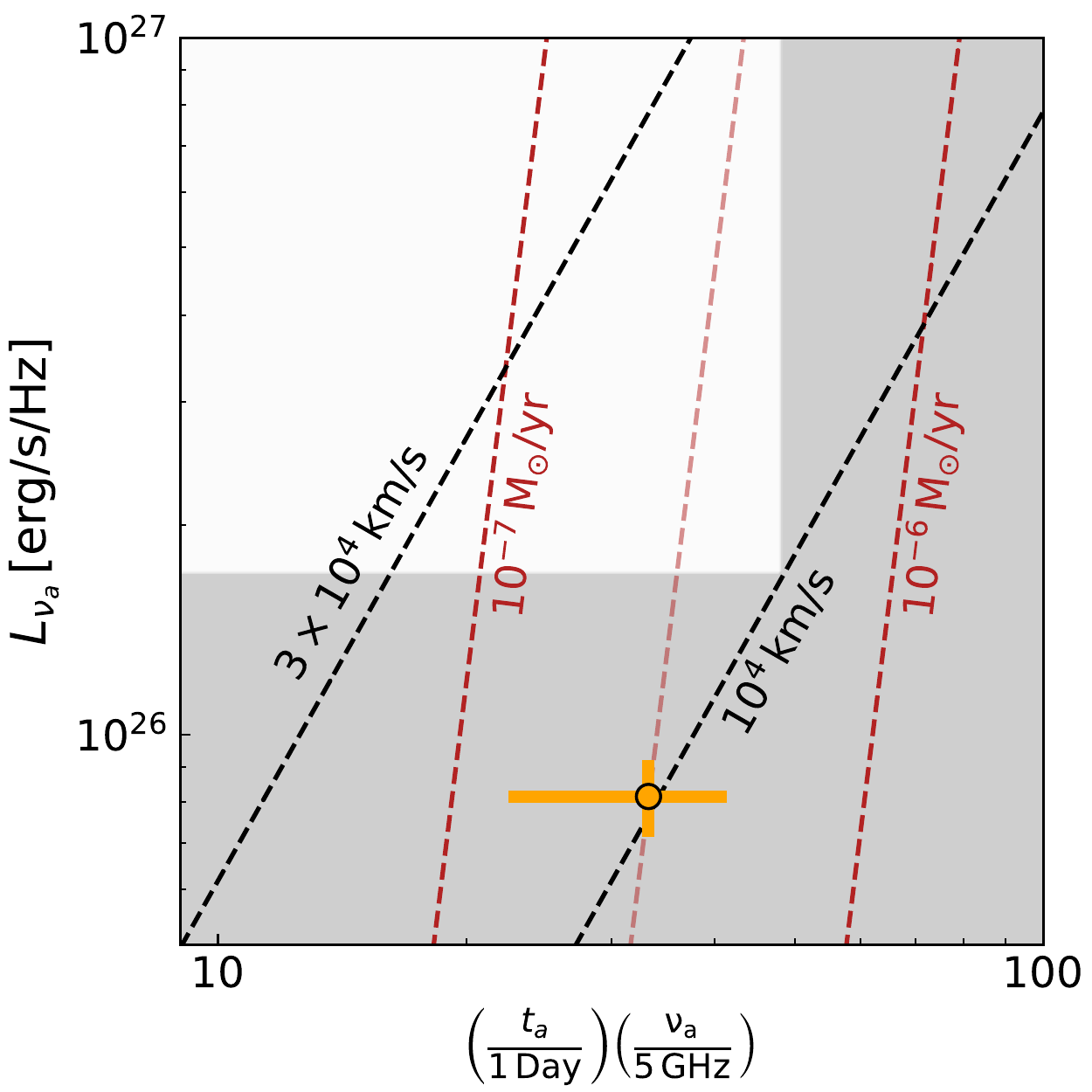}
    \caption{Chevalier's diagram for SN2020 qmp. The orange point is the position of the radio spectral peak derived $57$ days after the explosion. Under the assumption of a shockwave traveling with a constant velocity in a simple $\rm r^{-2}$ CSM density structure, the position of the peak should remain constant. However, the shaded region, which is the ruled out region derived from the limit on the radio spectral peak on the subsequent observation (on day $104$), is in disagreement with the radio spectral peak obtained before. This disagreement points toward a nontypical CSM structure. Also plotted for reference are equal lines of shock velocities and mass-loss rates (assuming a wind velocity of $10$ km/s).}
    \label{fig: Chevalier_diagram}
\end{figure}

The CSM interaction model with the assumptions presented above predicts a constant peak flux density that shifts to lower frequencies with time, for an assumed CSM structure of $\mathrm{r^{-2}}$. However, the peak flux density $104$ days after the explosion, despite not being observed, is higher than the observed peak flux density at $57$ days after the explosion. A useful tool for examining this atypical increase in the peak flux density between the two epochs is the phase space of peak radio spectral luminosity, $L_{\nu_a}$, versus the time of the peak, $t_a$, {multiplied} by its frequency, $\nu_a$. Lines of equal shock velocities and equal mass-loss rates can be plotted in this phase space, also known as Chevalier's diagram \citep{Chevalier1998}. Figure \ref{fig: Chevalier_diagram} shows Chevalier's diagram for SN\,2020qmp, with the radio emission spectral peak at $57$ days after explosion (see Sect. \ref{subsec: radio_modeling}). Also shown in this figure is a shaded region that marks the region ruled out due to the limit on the peak flux density and frequency $104$ days after the explosion. {Lines of equal shock velocities and mass-loss rates for values derived by the peak at 57 days are also plotted.} We assumed here a wind velocity of $10$ $\rm{km \, s^{-1}}$ and $p=3$.

Under the assumption of a constant shock velocity in the CSM and a CSM density structure of $r^{-2}$, the position of the peak in this phase space should remain constant over time. However, as seen in Fig. \ref{fig: Chevalier_diagram}, the peak flux density changes significantly between the two epochs. This suggests that (when assuming a constant shock velocity) the mass-loss rate varies significantly, by a factor of $\sim 2$. Furthermore, the lowest observed frequency on day 136 might feature a spectral turnover. If indeed that is the case, this points toward additional variability in the CSM density structure. However, we emphasize that since the possible turnover in the spectrum on day 136 is based only on one point, it should be treated with caution.

\subsection{Deviation from equipartition}
\label{equipartition}
The results of our radio emission modeling above are sensitive to the assumption of the ratio between the fraction of shock energy that goes into electron acceleration ($\epsilon_e$) and the fraction of energy that goes into the enhanced magnetic field ($\epsilon_B$). In our analysis above we used the common assumption of equipartition ($\alpha \equiv \epsilon_e / \epsilon_B = 1$). However, deviations from equipartition have been observed in several SNe (SN 2011dh \citep{soderberg_2012, Assaf2013}, SN 2012aw; \citep{yadav_2014}, SN 2013df \citep{Kamble_2016}, and SN 2020oi; \citep{horesh_2020}). Typically, when one has only radio data, it is difficult to determine whether this assumption holds. However, given we have an X-ray detection (albeit marginal), we can try to estimate these micro-physical parameters. We extrapolated the radio spectrum we observed at $57$ days after explosion to the time of X-ray detection ($9.6$ days) according to a typical power law for the optically thin regime of $F_{\nu} \left( t \right) \sim t^{-1}$ \citep{Chevalier1998}. We then extrapolated this emission to the X-ray band according to the spectral index obtained previously for that epoch. This gives an estimated luminosity of $2 \times 10^{36} \, \rm{erg \, s^{-1}}$ at the \emph{Swift}/XRT band on day $9.6$ after the explosion. This is three orders of magnitude lower than the observed X-ray luminosity at that time. Thus, even if the extrapolation of the radio emission to early times and to the X-ray band is somewhat crude, there is evidence for excess in X-ray emission.

\cite{bjornsson_2004} suggested inverse Compton (IC) scattering of photospheric photons by relativistic electrons at the shock front as a possible emission mechanism in the X-ray. Inverse Compton scattering is also assumed to be responsible for the observed X-ray emission in several past SNe (SN 2011dh \citealt{soderberg_2012, Assaf2013}, SN 2012aw; \citealt{yadav_2014}, SN 2013df \citealt{Kamble_2016}, and SN 2020oi; \citealt{horesh_2020}). Equation $32$ in \cite{Chevalier2006} gives the X-ray luminosity in the case of IC scattering as a function of the bolometric luminosity, the time after the explosion, the mass-loss rate, the shock velocity, and the microphysical parameters. We made use of this equation to estimate the mass-loss rate given the X-ray luminosity of $2\times 10^{39}$ $\rm{erg \, s^{-1}}$ $9.6$ days after the explosion, the bolometric luminosity of $2.1 \times 10^{42}$ $\rm{erg \, s^{-1}}$ at that time, and the assumption of equipartition. We assumed a shock velocity of $10^4$ $\rm{km \, s^{-1}}$ based on the optical photospheric expansion velocity of $\sim 9000$ $\rm{km \, s^{-1}}$ from optical spectra near that time. The value of the optically derived velocity is expected to be somewhat lower than the velocity of the shock in the CSM since the optical emission originates from a deeper and slower region of the SN ejecta. We infer a mass-loss rate of $1.04 \times 10^{-5} \, \rm M_{\odot}/yr$ for an assumed wind velocity of $10$ $\rm{km \, s^{-1}}$. On the other hand, the mass-loss rate derived from the radio spectrum $57$ days after the explosion is smaller by a factor of $\sim 30$ (see Sect. 5.1). 

As seen above, under the assumption of equipartition, the mass-loss rate derived from the radio synchrotron emission on day $57$ is in disagreement with the mass-loss rate derived from the X-ray IC emission on day $9.6$. A possible explanation for this large discrepancy in the mass-loss rate estimates is a deviation from equipartition, such that it satisfies $\epsilon_B = 2 \times 10^{-4}$, which in turn translates to $\alpha = 484$ (assuming a typical $\epsilon_e = 0.1$). This, in turn, results in a reduction in the shockwave radius estimate (and in the shock velocity estimate) derived from the radio spectrum on day $57$ of $28\%$. The shock velocity estimate decreases to $\mathrm{v_{sh}}=7200$ $\rm{km \, s^{-1}}$ in this case. Alternatively, the above discrepancy can be reconciled if the progenitor star experiences huge variability in its mass-loss rate (by a factor of $\sim 30$ at least) in the years before the explosion. While the radio data alone suggest some variability in the mass-loss rate, this variability is only a factor of $\sim 2$. The level of this observed mass-loss rate variability is far from the variability needed to explain the observed X-ray emission under the equipartition assumption. We emphasize that since the X-ray detection on day $9.6$ is only at the level of $2 \sigma$, any physical parameter inferred from it (i.e., mass-loss rate, shock velocity, and the microphysical parameters) should be taken with a grain of salt.

\section{Local CCSN rate and near-infrared surveys} 
\label{rate}
Most massive star formation occurs in highly dust-obscured regions in the Universe, and consequently this is where most CCSNe are found. Therefore, determining the rate of CCSNe is highly dependent on the effects of dust and extinction \citep[][]{Grossan1999, Maiolino2002, Miluzio2013, Kool2017, SpitzerDust2021}. As a NIR survey, PGIR is sensitive to CCSNe that may be obscured at optical wavelengths due to high extinction values. Though SN 2020qmp itself does not appear to be highly extinguished at optical wavelengths, 
its discovery begs the question as to how effective NIR surveys are in detecting obscured SNe in comparison to optical surveys.
 
\citet{Mattila2012} derived a CCSN rate of $7.4^{+3.7}_{-2.6} \times 10^{-4} \, \mathrm{yr}^{-1} \, \mathrm{Mpc}^{-3}$ within the local 6 Mpc volume and $1.5^{+0.4}_{-0.3} \times 10^{-4} \, \mathrm{yr}^{-1} \,  \mathrm{Mpc}^{-3}$ within the local 6-15 Mpc volume, using a 12 year sample of CCSNe from 2000-2012. However, they also derived an estimate of $18.9^{+19.2}_{-9.5} \%$ of CCSNe missed locally by optical surveys. More recently, \citet{Jencson2019} found that this number could be as high as $38.5^{+26.0}_{-21.9} \%$, {though defined in a somewhat different way by not taking into account the effects of host galaxy inclination}, based on highly reddened CCSNe detected in a sample of nearby galaxies ($D\lesssim40$\,Mpc) in the mid-infrared by the Spitzer Infrared Intensive Transient Survey (SPIRITS; \citealp{Kasliwal2017}). As some objects in the sample were not definitively classified as CCSNe, we consider this as a maximal estimate of the optically missed fraction. Here, we examine the sensitivity of wide-field, ground-based surveys in the NIR -- such as PGIR and the future Wide-Field Infrared Transient Explorer (WINTER) survey, an upcoming (first light planned for Fall 2021) $J$-band search with a  1-meter telescope at Palomar Observatory that is expected to achieve a median depth of 20.8 AB mag \citep{Simcoe2019, Lourie2021, Frostig2021} over the entire accessible northern sky at a cadence of a few weeks -- to detecting such highly obscured CCSNe in the local Universe. {We note that if there is a bright background galaxy associated with a CCSN, then the expected median depth will be lower than the numbers reported. PGIR will be more affected than WINTER by bright background galaxies as the former has an 8 arcsecond pixel scale and the latter a 1 arcsecond pixel scale. Therefore, WINTER's median depth will only be significantly affected by the small fraction of transients that occur deep in nuclear regions of galaxies.} 

\begin{figure}
    \centering
      \includegraphics[width=\linewidth]{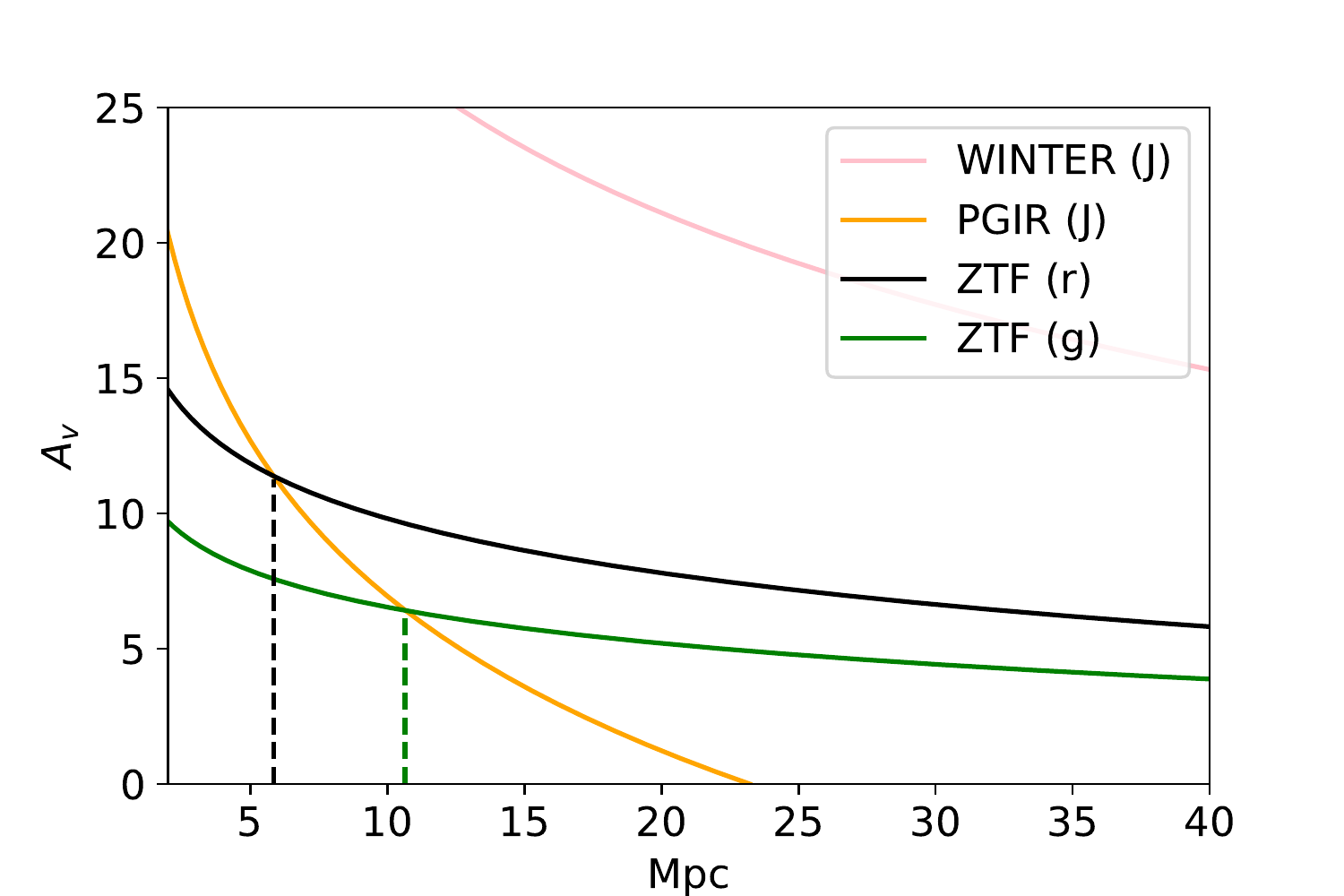}
       \includegraphics[width=\linewidth]{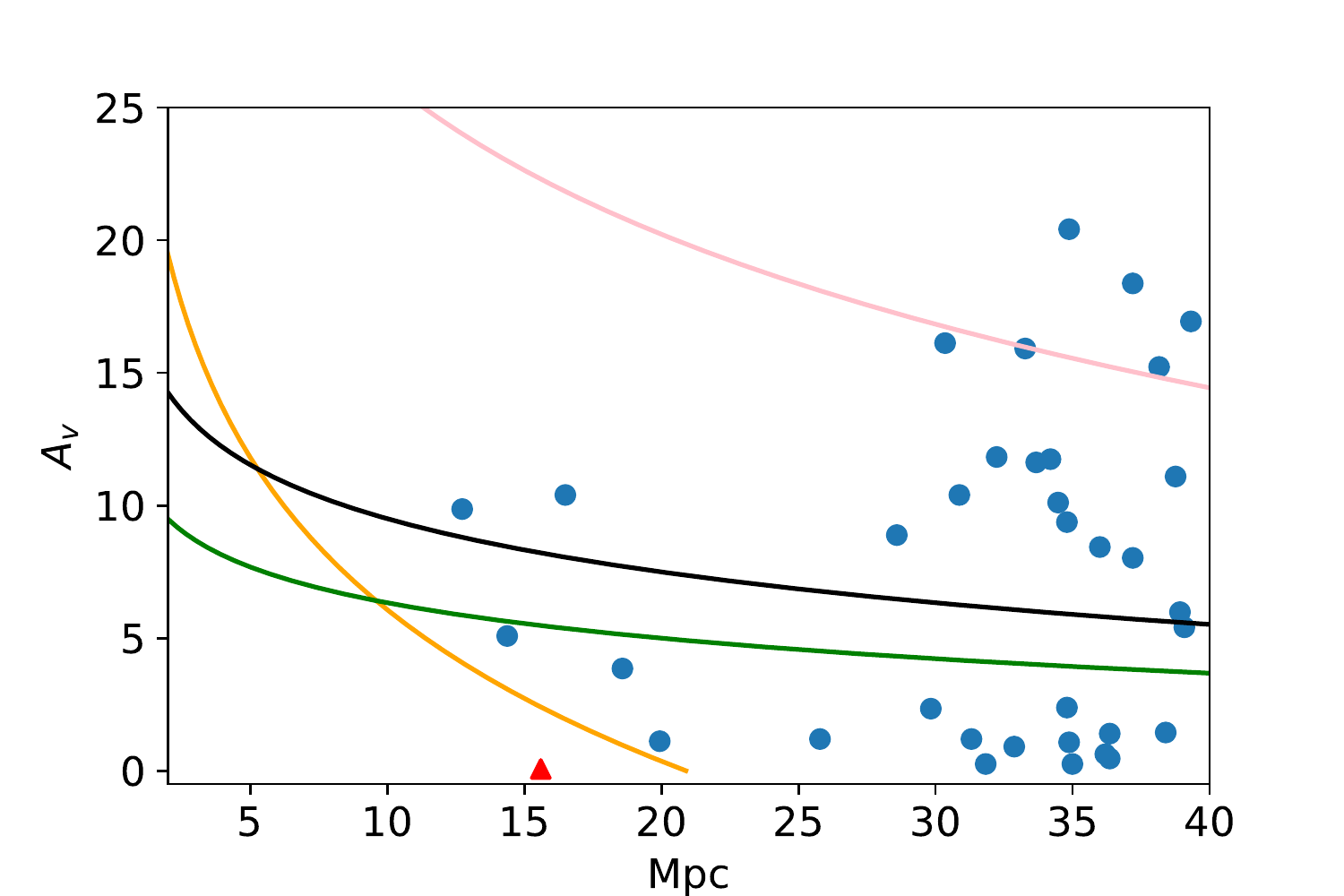}
    \caption{Sensitivity Curves and CCSNe simulation detailed in the text. \textit{Top panel:} Comparison of PGIR's sensitivity to detecting {Type IIP CCSNe} as a function of distance and extinction to that of ZTF's $r$ and $g$ bands, assuming an absolute AB magnitude of -16.8 for every SN. The proposed WINTER survey's sensitivity is also shown. The vertical lines are the distances (in Mpc) where PGIR is more sensitive to extinguished CCSNe than ZTF. \textit{Bottom panel:} Simulated {Type IIP CCSNe} over a five-year span using an extinction distribution derived from the CCSN candidate sample of \citet{Jencson2019}. Details of the simulation are given in the main text, and the sensitivity curves are the same as in the top panel. SN 2020qmp in particular is marked in red and as a triangle to demonstrate its placement compared to the simulated distribution.}
    \label{Sncompare}
\end{figure}

\citet{Richardson2014} calculated the bias-corrected absolute magnitude distributions of SNe primarily from the Asiago Supernova Catalog \citep{Asiago} as well as a few supplemental data sources. {They find the average absolute magnitude of Type IIP SNe in their volume-limited sample to be $M_{AB} = -16.8 \pm 0.37$.} Assuming this absolute magnitude for discovery, in the upper panel of Fig. \ref{Sncompare} we show the sensitivity of PGIR in the $J$ band to detecting {Type IIP CCSNe} at a given distance and value of the total $V$-band extinction, $A_V$, assuming a survey depth of $J = 15.7$ mag. To convert from $A_V$ to $A_J$, the extinction value in the $J$ band, we assumed a standard extinction law with $R_V = 3.1$ according to \citet{Fitzpatrick} and used a 10000 K blackbody source spectrum as an approximation to a CCSN near peak light, and we{ derived a ratio of $A_J/A_V = 0.264$}. We also show the corresponding sensitivity curves for ZTF in the $g$ and $r$ bands, assuming a survey depth of 20.5 mag for both bands. 
Figure \ref{Sncompare} shows that PGIR is more effective in detecting very highly obscured Type IIP CCSNe ($A_V \gtrsim 10$--$15$\,mag) in the very local Universe despite the fact that its median depth is $\sim$ 5 mag {shallower} than ZTF. 
Specifically, using the formulation detailed above, PGIR is more sensitive to these extinguished CCSNe than ZTF out to $\sim$ {5.3} Mpc in the $r$ band and $\sim$ {9.6} Mpc in the $g$ band. We also show the vast improvement in sensitivity to obscured CCSNe for WINTER, assuming a median survey depth of $J = 20.8$ mag. We note that we do not take the cadence of WINTER  into account, assuming it is sufficient to catch any SN in its active footprint near peak light. {We repeated this analysis for the different subclasses of CCSNe (Type Ib/c, IIb, IIP, IIL, and IIn) and report our results in Table \ref{SNcomparetablereal}.}

\begin{centering}
\begin{table*}
\begin{tabular}{||c c c c c c||}
\hline
{CCSN Type} & AB mag & {PGIR $\sim$ ZTF $r$ (Mpc)} & {PGIR $\sim$ ZTF $g$ (Mpc)} & {N Expected} & {N Detected by WINTER $\sim$ ZTF}  \\
\hline
Ib/c & -17.61 & 7.6 & 13.8 & 19 & 3.7\\
IIb & -17.03 & 5.9 & 10.6 & 7 & 1.3 \\
IIP & -16.8 & 5.3 & 9.6 & 39 & 7.3 \\
IIL & -17.98 & 9.1 & 16.5 & 5 & 1.0 \\
IIn & -18.62 & 12.2 & 22.1 & 5 & 1.0\\
\hline
\end{tabular}
\caption{{Results of the CCSN simulation run for every subclass of CCSNe. We report: the average absolute AB magnitude; the distance in Mpc  where PGIR is more sensitive to extinguished CCSNe when compared to ZTF's $r$ band and $g$ band; the number of expected of events in a five-year time period using the CCSN rate from \citet{Mattila2012} and subclass fractions from \cite{Ratepaper}; and the number of events we expect WINTER to detect that ZTF misses. }}
\label{SNcomparetablereal}
\end{table*}
\end{centering}

\begin{figure}
    \centering
    \includegraphics[width=\linewidth]{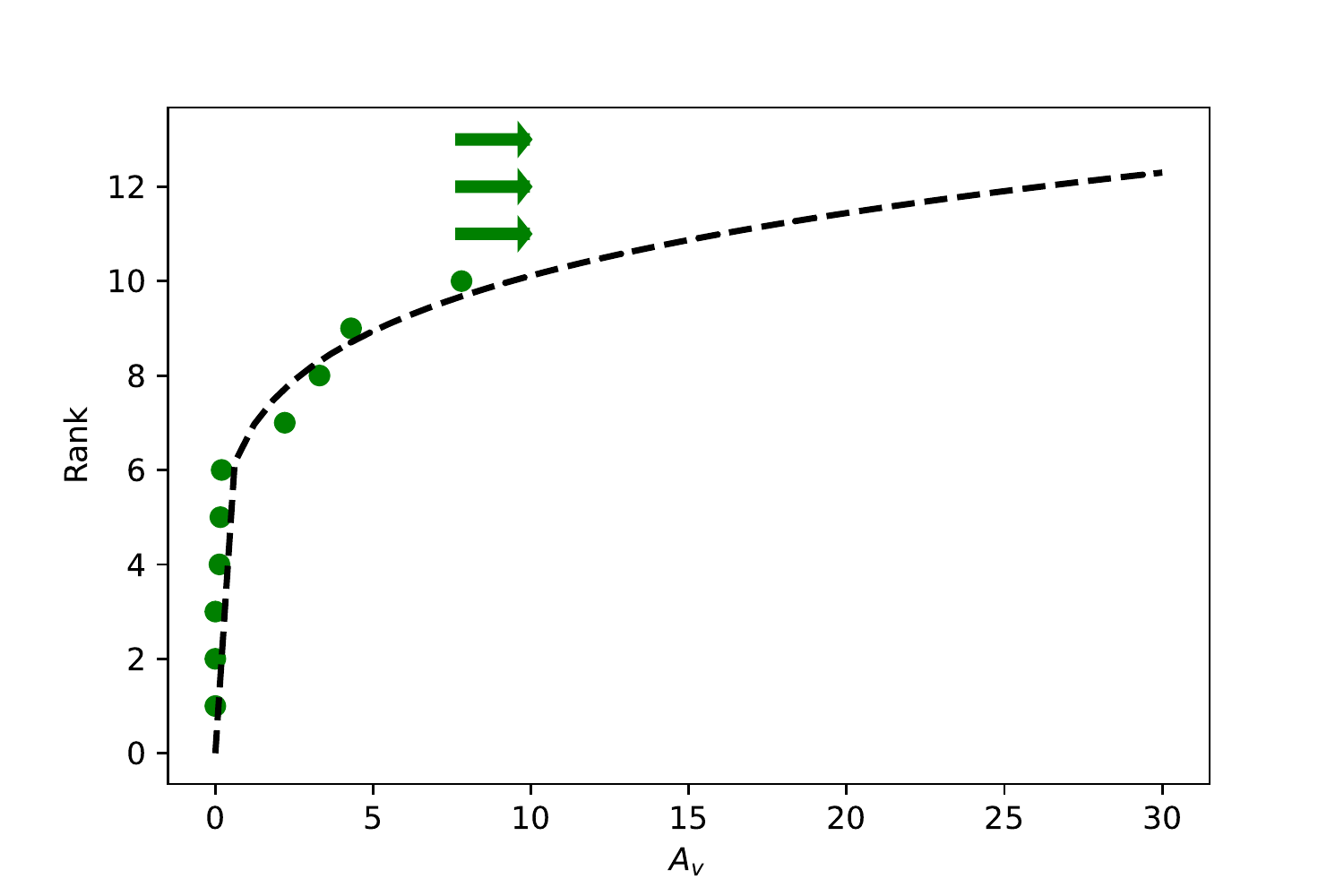}
    \caption{{Cumulative distribution along with a power-law fit of the extinction distribution from \citet{Jencson2019} used in the simulation. This cumulative distribution represents the most extreme extinction scenario, with all lower limits in the sample assigned the highest extinction value in the sample of $A_V = 7.8$, indicated with arrows.}}
    \label{JacobExtinction}
\end{figure}

We then ran a simulation representing a distribution of CCSNe in the local 40\,Mpc volume with varying levels of extinction. In order to model the extinction distribution, we pull from \citet{Jencson2019}, using their sample of optically {discovered} CCSNe (eight) and infrared-discovered confirmed (two) and candidate CCSNe (two) in the SPIRITS sample between 2014 and 2018. We constructed an empirical extinction distribution based on the reported values of $A_V$ for each object. Three of the infrared-discovered objects were undetected in the optical, and thus only lower limits on $A_V$ for these sources were available. We assigned them values $A_V = 7.8$\,mag in the distribution, the highest measured value of any object in the sample, and emphasize again that we consider this as a {maximal} estimate of the fraction of highly reddened CCSNe. {Therefore, the results we obtain from our simulation represent the {upper limit} on the number of CCSNe that can be detected by PGIR and WINTER with respect to ZTF, and they represent the most extreme extinction scenario.} {The cumulative distribution is shown in Fig. \ref{JacobExtinction}}.

We then fitted a power law to the cumulative distribution of measured $A_V$ values and from there derived a probability distribution function for the extinction values of CCSNe. Next, we derived a probability distribution for the distances of CCSNe that are uniformly distributed throughout the local 40\,Mpc volume, where we adopted the volumetric CCSN rate of \citet{Mattila2012} mentioned above for $D<6$\,Mpc and applied the 6--15 Mpc value for larger distances. For an assumed active areal survey coverage of 15,000~sq.~deg (approximately the entire accessible northern sky from Palomar at any given time), we thus expect 75 total CCSNe within the 40\,Mpc volume in a five-year period to fall within the active survey footprint of PGIR, ZTF, or WINTER. {We then divided these 75 CCSNe into different subclasses using the observed fractions of subclasses that \citet{Ratepaper} derived from a  volume-limited sample of 60 Mpc, and we assigned the same absolute magnitudes for each subclass used earlier from \citet{Richardson2014}. We report the number of expected events during a five-year time period for each subclass in Table \ref{SNcomparetablereal}}.

Finally, we ran the simulation {separately for each subclass by distributing the number of events obtained} randomly in distance and $A_V$ according to the probability distributions derived above, and 
and repeated the simulation 1000 times. A single instance of the simulation run for the Type IIP class is shown in Fig. \ref{Sncompare}. We caution that the spread of CCSNe at high extinction values above $A_V \gtrsim 8$ mag is based on an extrapolation of the empirical distribution derived from the SPIRITS sample and should be viewed with caution. Events that fall below a given sensitivity curve are counted as ``detected'' by the respective survey. The results for every subclass are reported in Table \ref{SNcomparetablereal}.

On average, over the 1000 simulations, we find that PGIR is not expected to detect any CCSNe that ZTF would miss due to dust extinction in a five-year time period. 
{This} is mainly attributable to the fact that PGIR's median depth (15.7 AB mag) is much lower than that of ZTF (20.5 AB mag). However, when looking at future NIR surveys such as WINTER, we find a much higher number of projected CCSNe that would be missed by ZTF but are accessible to WINTER.  {Adding the results over every subclass}, we expect a total of around {14} CCSNe ($\approx$18\% of the 75 total) accessible to WINTER that would be missed by ZTF. WINTER's median depth (20.8 AB mag) is around the same as ZTF's and clearly demonstrates the advantage that employing NIR surveys can have on discovering highly reddened CCSNe in the future. While this simulation is based on the assumption of a maximal estimate of the number of highly reddened SNe, it demonstrates that a five-year SN search with a deep, wide-field survey with WINTER will have sufficient number statistics to accurately constrain the high end of the extinction distribution for CCSNe in the local Universe. 

\section{Conclusions}
\label{Conclusions}
In this paper we present a detailed, multiwavelength analysis of SN 2020qmp, which was discovered by PGIR. Based on characteristic hydrogen lines in its spectra, along with a long plateau in its optical LC, the SN can be classified as a Type IIP SN. We do not detect any polarization from the SN during the plateau phase, which is expected because the outer ejecta visible during this phase {are} generally symmetric \citep{Wang2008}.  By comparing the normalized line luminosities of the [O I] doublet (relative to the $^{56} \mathrm{Co}$ decay energy) between the observed spectrum and \citet{Jerkstrand2014} models, we estimate the ZAMS progenitor mass of the SN to be {between $9$ and $12 \, \mathrm{M_\odot}$. Through hydrodynamical LC modeling, we find the explosion energy to be around $0.8 \times 10^{51} \mathrm{erg}$ and better constrain the ZAMS progenitor mass to around $11.0 \, \mathrm{M_\odot}$, which are values comparable to those obtained in analyses done of other Type IIP progenitors \citep{Anderson2014, Sanders_2015}}

We also make use of broadband radio observations conducted with the VLA to derive the physical properties of the shock in the CSM under the CSM interaction model. Assuming equipartition between the fraction of energy in the electron ($\epsilon_e$) and the fraction of energy in the enhanced magnetic field ($\epsilon_B$), the radio spectrum $57$ days after the explosion gives a shock velocity of ${v_{sh}} = 10^4$ $\rm{km \, s^{-1}}$ and a mass-loss rate of $\mathrm{\dot{M}} = \left( 2.9^{+1.0} _{-1.2} \right) \times 10^{-7} \, \rm M_{\odot} \, yr^{-1}$, for an assumed wind velocity of $10$ $\rm{km \, s^{-1}}$, which is within the range of most red supergiant progenitors for Type IIP SNe \citep{radioreview}. However, the radio spectrum on day $104$ showed a surprisingly higher peak flux density at lower frequency than the one observed on day $57$. We determine that, assuming standard CSM interaction models and a constant shock velocity, an increase in the mass-loss rate by a factor of $\sim 2$ is needed to explain this discrepancy. This and additional radio observations on day $136$ point to variability in the progenitor mass-loss rate during the $1000$\,years prior to explosion.

Early X-ray observations with Swift/XRT show tentative excess emission compared to observations extrapolated to the same epoch from radio frequencies using a standard shockwave evolution. Assuming that this emission excess originates from IC scattering of {photospheric} photons by relativistic electrons in the shock front, we derive a much greater mass-loss rate than the one derived by the radio spectrum on day $57$, of $\mathrm{\dot{M}} = 1.04 \times 10^{-5} \rm \, {M_{\odot} \, yr^{-1}}$, for an assumed wind velocity of $10$ $\rm{km \, s^{-1}}$. This discrepancy can be resolved assuming deviation from equipartition, $\epsilon_B = 0.0002$, and $\epsilon_e = 0.1$. This also calls for a reduction in the inferred shock velocity of $28\%$, from $10^4$ $\rm{km \, s^{-1}}$ to $7200$ $\rm{km \, s^{-1}}$. {Alternatively, one can also explain the difference in mass-loss rates} via  the extreme mass-loss variations from the progenitor in the years prior to the explosion.

Finally, we created a simulation of CCSNe within a five-year span assuming CCSN rates from \citet{Mattila2012}, extrapolating an extinction distribution from \citet{Jencson2019}, {representing a maximal estimate of the fraction of highly reddened CCSNe},  and {assuming  absolute magnitudes for different subclasses of CCSNe from \citet{Richardson2014}}. Though SN 2020qmp itself is not extremely extinguished, with $A_v = 0.0669$ \citep{Schlafly2011}, its discovery prompted the question as to how much more sensitive PGIR is to extinguished SNe as a NIR survey relative to optical surveys such as ZTF. We find that in a five-year span, we do not expect PGIR to detect any CCSNe that ZTF misses, and this is due to the extremely lower median depth that PGIR has (15.7 AB mag) in comparison to ZTF (20.8 AB mag). However, this number shoots up when looking at future NIR surveys such as WINTER, which has higher median depths (21 AB mag in the $J$ band); we estimate  that WINTER will discover around 14 CCSNe that are missed by ZTF. This shows how promising future NIR surveys will be for discovering extinguished CCSNe. 

\small\section{Acknowledgements}
Palomar Gattini-IR (PGIR) is generously funded by Caltech, Australian National University, the Mt Cuba Foundation, the Heising Simons Foundation, the Binational Science Foundation. PGIR is a collaborative project among Caltech, Australian National University, University of New South Wales, Columbia University and the Weizmann Institute of Science. MMK acknowledges generous support from the David and Lucille Packard Foundation. MMK and EO acknowledge the US-Israel Bi-national Science Foundation Grant 2016227. MMK and JLS acknowledge the Heising-Simons foundation for support via a Scialog fellowship of the Research Corporation. MMK and AMM acknowledge the Mt Cuba foundation. J. Soon is supported by an Australian Government Research Training Program (RTP) Scholarship. A.H. acknowledges support by the I-Core Program of the Planning and Budgeting Committee and the Israel Science Foundation, and support by ISF grant 647/18.This research was supported by Grant No. 2018154 from the United States-Israel Binational Science Foundation (BSF). We thank the National Radio Astronomy Observatory (NRAO) for conducting the radio observations with the Karl G. Jansky Very Large Array (VLA). Some of the data presented here were obtained with the Visiting Astronomer facility at the Infrared Telescope Facility, which is operated by the University of Hawaii under contract 80HQTR19D0030 with the National Aeronautics and Space Administration. Some of the data presented herein were obtained at the W.M. Keck Observatory, which is operated as a scientific partnership among the California Institute of Technology, the University of California and the National Aeronau- tics and Space Administration. The Observatory was made possible by the generous financial support of the W.M. Keck Foundation. SED Machine is based upon work supported by the National Science Foundation under Grant No. 1106171. The authors wish to recognize and acknowledge the very significant cultural role and reverence that the summit of Mauna Kea has always
had within the indigenous Hawaiian community. We are most fortunate to have the opportunity to conduct observations from this mountain. Based on observations obtained with the Samuel Oschin Telescope 48-inch and the 60-inch Telescope at the Palomar Observatory as part of the Zwicky Transient Facility project. ZTF is supported by the National Science Foundation under Grant No. AST-1440341 and a collaboration including Caltech, IPAC, the Weizmann Institute for Science, the Oskar Klein Center at Stockholm University, the University of Maryland, the University of Washington, Deutsches Elektronen-Synchrotron and Humboldt University, Los Alamos National Laboratories, the TANGO Consortium of Taiwan, the University of Wisconsin at Milwaukee, and Lawrence Berkeley National Laboratories. Operations are conducted by COO, IPAC, and UW. We thank Yize Dong for his help with the hydrodynamical LC modeling. We thank David Kaplan for his helpful comments before submission to the journal. 

\bibliographystyle{aa}
\bibliography{PGIRNRNAAS.tex}

\begin{thebibliography}{106}
\expandafter\ifx\csname natexlab\endcsname\relax\def\natexlab#1{#1}\fi

\bibitem[{Anderson {et~al.}(2014)Anderson, González-Gaitán, Hamuy,
  Gutiérrez, Stritzinger, Olivares~E., Phillips, Schulze, Antezana, Bolt, \&
  et~al.}]{Anderson2014}
Anderson, J.~P., González-Gaitán, S., Hamuy, M., {et~al.} 2014, \apj, 786, 67

\bibitem[{{Arnett}(1980)}]{Arnett1980}
{Arnett}, W.~D. 1980, \apj, 237, 541

\bibitem[{{Bellm} {et~al.}(2019){Bellm}, {Kulkarni}, {Graham}, {Dekany},
  {Smith}, {Riddle}, {Masci}, {Helou}, {Prince}, {Adams}, {Barbarino},
  {Barlow}, {Bauer}, {Beck}, {Belicki}, {Biswas}, {Blagorodnova}, {Bodewits},
  {Bolin}, {Brinnel}, {Brooke}, {Bue}, {Bulla}, {Burruss}, {Cenko}, {Chang},
  {Connolly}, {Coughlin}, {Cromer}, {Cunningham}, {De}, {Delacroix}, {Desai},
  {Duev}, {Eadie}, {Farnham}, {Feeney}, {Feindt}, {Flynn}, {Franckowiak},
  {Frederick}, {Fremling}, {Gal-Yam}, {Gezari}, {Giomi}, {Goldstein},
  {Golkhou}, {Goobar}, {Groom}, {Hacopians}, {Hale}, {Henning}, {Ho}, {Hover},
  {Howell}, {Hung}, {Huppenkothen}, {Imel}, {Ip}, {Ivezi{\'c}}, {Jackson},
  {Jones}, {Juric}, {Kasliwal}, {Kaspi}, {Kaye}, {Kelley}, {Kowalski},
  {Kramer}, {Kupfer}, {Landry}, {Laher}, {Lee}, {Lin}, {Lin}, {Lunnan},
  {Giomi}, {Mahabal}, {Mao}, {Miller}, {Monkewitz}, {Murphy}, {Ngeow},
  {Nordin}, {Nugent}, {Ofek}, {Patterson}, {Penprase}, {Porter}, {Rauch},
  {Rebbapragada}, {Reiley}, {Rigault}, {Rodriguez}, {van Roestel}, {Rusholme},
  {van Santen}, {Schulze}, {Shupe}, {Singer}, {Soumagnac}, {Stein}, {Surace},
  {Sollerman}, {Szkody}, {Taddia}, {Terek}, {Van Sistine}, {van Velzen},
  {Vestrand}, {Walters}, {Ward}, {Ye}, {Yu}, {Yan}, \& {Zolkower}}]{Bellm2019}
{Bellm}, E.~C., {Kulkarni}, S.~R., {Graham}, M.~J., {et~al.} 2019, \pasp, 131,
  018002

\bibitem[{{Bellm} \& {Sesar}(2016)}]{Bellm2016}
{Bellm}, E.~C. \& {Sesar}, B. 2016, {pyraf-dbsp: Reduction pipeline for the
  Palomar Double Beam Spectrograph}

\bibitem[{{Ben-Ami} {et~al.}(2012){Ben-Ami}, {Gal-Yam}, {Filippenko},
  {Mazzali}, {Modjaz}, {Yaron}, {Arcavi}, {Cenko}, {Horesh}, {Howell},
  {Graham}, {Horst}, {Im}, {Jeon}, {Kulkarni}, {Leonard}, {Perley}, {Pian},
  {Sand}, {Sullivan}, {Becker}, {Bersier}, {Bloom}, {Bottom}, {Brown}, {Clubb},
  {Dilday}, {Dixon}, {Fortinsky}, {Fox}, {Gonzalez}, {Harutyunyan}, {Kasliwal},
  {Li}, {Malkan}, {Manulis}, {Matheson}, {Moskovitz}, {Muirhead}, {Nugent},
  {Ofek}, {Quimby}, {Richards}, {Ross}, {Searcy}, {Silverman}, {Smith},
  {Vanderburg}, \& {Walker}}]{BenAmi12}
{Ben-Ami}, S., {Gal-Yam}, A., {Filippenko}, A.~V., {et~al.} 2012, \apjl, 760,
  L33

\bibitem[{{Berger} {et~al.}(2002){Berger}, {Kulkarni}, \&
  {Chevalier}}]{Berger2002}
{Berger}, E., {Kulkarni}, S.~R., \& {Chevalier}, R.~A. 2002, \apjl, 577, L5

\bibitem[{{Bietenholz} {et~al.}(2021){Bietenholz}, {Bartel}, {Argo}, {Dua},
  {Ryder}, \& {Soderberg}}]{Bietenholtz}
{Bietenholz}, M.~F., {Bartel}, N., {Argo}, M., {et~al.} 2021, \apj, 908, 75

\bibitem[{{Bj{\"o}rnsson} \& {Fransson}(2004)}]{bjornsson_2004}
{Bj{\"o}rnsson}, C.-I. \& {Fransson}, C. 2004, \apj, 605, 823

\bibitem[{Blagorodnova {et~al.}(2018)Blagorodnova, Neill, Walters, Kulkarni,
  Fremling, Ben-Ami, Dekany, Fucik, Konidaris, Nash, Ngeow, Ofek, Sullivan,
  Quimby, Ritter, \& Vyhmeister}]{Blagorodnova_2018}
Blagorodnova, N., Neill, J.~D., Walters, R., {et~al.} 2018, \pasp, 130, 035003

\bibitem[{Branch \& Wheeler(2017)}]{Branch2017}
Branch, D. \& Wheeler, J.~C. 2017, Type IIP Supernovae (Berlin, Heidelberg:
  Springer Berlin Heidelberg), 245--265

\bibitem[{{Breeveld} {et~al.}(2011){Breeveld}, {Landsman}, {Holland}, {Roming},
  {Kuin}, \& {Page}}]{Breeveld2011a}
{Breeveld}, A.~A., {Landsman}, W., {Holland}, S.~T., {et~al.} 2011, in American
  Institute of Physics Conference Series, Vol. 1358, American Institute of
  Physics Conference Series, ed. J.~E. {McEnery}, J.~L. {Racusin}, \&
  N.~{Gehrels}, 373--376

\bibitem[{{Burrows} {et~al.}(2005){Burrows}, {Hill}, {Nousek}, {Kennea},
  {Wells}, {Osborne}, {Abbey}, {Beardmore}, {Mukerjee}, {Short}, {Chincarini},
  {Campana}, {Citterio}, {Moretti}, {Pagani}, {Tagliaferri}, {Giommi},
  {Capalbi}, {Tamburelli}, {Angelini}, {Cusumano}, {Br{\"a}uninger}, {Burkert},
  \& {Hartner}}]{Burrows2005a}
{Burrows}, D.~N., {Hill}, J.~E., {Nousek}, J.~A., {et~al.} 2005, \ssr, 120, 165

\bibitem[{{Chevalier}(1982)}]{Chevalier1982}
{Chevalier}, R.~A. 1982, \apj, 259, 302

\bibitem[{{Chevalier}(1998)}]{Chevalier1998}
{Chevalier}, R.~A. 1998, \apj, 499, 810

\bibitem[{{Chevalier} \& {Fransson}(2006)}]{Chevalier2006}
{Chevalier}, R.~A. \& {Fransson}, C. 2006, \apj, 651, 381

\bibitem[{{Chonis} \& {Gaskell}(2008)}]{Chonis2008}
{Chonis}, T.~S. \& {Gaskell}, C.~M. 2008, \aj, 135, 264

\bibitem[{{Cushing} {et~al.}(2004){Cushing}, {Vacca}, \&
  {Rayner}}]{Cushing2004}
{Cushing}, M.~C., {Vacca}, W.~D., \& {Rayner}, J.~T. 2004, \pasp, 116, 362

\bibitem[{{De} {et~al.}(2020{\natexlab{a}}){De}, {Hankins}, {Kasliwal},
  {Moore}, {Ofek}, {Ofek}, {Ofek}, {Ofek}, {Ofek}, \& {Ofek}}]{de2020}
{De}, K., {Hankins}, M., {Kasliwal}, M., {et~al.} 2020{\natexlab{a}}, \pasp,
  132, 025001

\bibitem[{{De} {et~al.}(2020{\natexlab{b}}){De}, {Hankins}, {Kasliwal},
  {Ashley}, {Babul}, {Jencson}, {Karambelkar}, {Lau}, {Moore}, {Ofek},
  {Sharma}, {Sokoloski}, {Soon}, {Soria}, \& {Travouillon}}]{ATEL}
{De}, K., {Hankins}, M., {Kasliwal}, M.~M., {et~al.} 2020{\natexlab{b}}, The
  Astronomer's Telegram, 13909, 1

\bibitem[{{De} {et~al.}(2021){De}, {Kasliwal}, {Hankins}, {Sokoloski}, {Adams},
  {Ashley}, {Babul}, {Bagdasaryan}, {Delacroix}, {Dekany}, {Greffe}, {Hale},
  {Jencson}, {Karambelkar}, {Lau}, {Mahabal}, {McKenna}, {Moore}, {Ofek},
  {Sharma}, {Smith}, {Soon}, {Soria}, {Srinivasaragavan}, {Tinyanont},
  {Travouillon}, {Tzanidakis}, \& {Yao}}]{De2021}
{De}, K., {Kasliwal}, M.~M., {Hankins}, M.~J., {et~al.} 2021, \apj, 912, 19

\bibitem[{{Dekany} {et~al.}(2020){Dekany}, {Smith}, {Riddle}, {Feeney},
  {Porter}, {Hale}, {Zolkower}, {Belicki}, {Kaye}, {Henning}, {Walters},
  {Cromer}, {Delacroix}, {Rodriguez}, {Reiley}, {Mao}, {Hover}, {Murphy},
  {Burruss}, {Baker}, {Kowalski}, {Reif}, {Mueller}, {Bellm}, {Graham}, \&
  {Kulkarni}}]{Dekaney2020}
{Dekany}, R., {Smith}, R.~M., {Riddle}, R., {et~al.} 2020, \pasp, 132, 038001

\bibitem[{{Dessart} {et~al.}(2010){Dessart}, {Livne}, \&
  {Waldman}}]{Dessart2010}
{Dessart}, L., {Livne}, E., \& {Waldman}, R. 2010, \mnras, 408, 827

\bibitem[{{Evans} {et~al.}(2009){Evans}, {Beardmore}, {Page}, {Osborne},
  {O'Brien}, {Willingale}, {Starling}, {Burrows}, {Godet}, {Vetere}, {Racusin},
  {Goad}, {Wiersema}, {Angelini}, {Capalbi}, {Chincarini}, {Gehrels}, {Kennea},
  {Margutti}, {Morris}, {Mountford}, {Pagani}, {Perri}, {Romano}, \&
  {Tanvir}}]{Evans2009a}
{Evans}, P.~A., {Beardmore}, A.~P., {Page}, K.~L., {et~al.} 2009, \mnras, 397,
  1177

\bibitem[{{Evans} {et~al.}(2007){Evans}, {Beardmore}, {Page}, {Tyler},
  {Osborne}, {Goad}, {O'Brien}, {Vetere}, {Racusin}, {Morris}, {Burrows},
  {Capalbi}, {Perri}, {Gehrels}, \& {Romano}}]{Evans2007a}
{Evans}, P.~A., {Beardmore}, A.~P., {Page}, K.~L., {et~al.} 2007, \aap, 469,
  379

\bibitem[{Filippenko(1997)}]{Filippenko1997}
Filippenko, A.~V. 1997, Annual Review of Astronomy and Astrophysics, 35, 309

\bibitem[{{Fitzpatrick}(1999)}]{Fitzpatrick}
{Fitzpatrick}, E.~L. 1999, \pasp, 111, 63

\bibitem[{Foreman-Mackey {et~al.}(2013)Foreman-Mackey, Hogg, Lang, \&
  Goodman}]{Foreman_Mackey_2013}
Foreman-Mackey, D., Hogg, D.~W., Lang, D., \& Goodman, J. 2013, \pasp, 125, 306

\bibitem[{Fox {et~al.}(2021)Fox, Khandrika, Rubin, Casper, Li, Szalai, Armus,
  Filippenko, Skrutskie, Strolger, \& Van Dyk}]{SpitzerDust2021}
Fox, O.~D., Khandrika, H., Rubin, D., {et~al.} 2021, Monthly Notices of the
  Royal Astronomical Society, 506, 4199

\bibitem[{{Frostig} {et~al.}(2020){Frostig}, {Baker}, {Brown}, {Burruss},
  {Clark}, {F{\.z}r{\'e}sz}, {Ganciu}, {Hinrichsen}, {Karambelkar}, {Kasliwal},
  {Lourie}, {Malonis}, {Simcoe}, \& {Zolkower}}]{Frostig2021}
{Frostig}, D., {Baker}, J.~W., {Brown}, J., {et~al.} 2020, in Society of
  Photo-Optical Instrumentation Engineers (SPIE) Conference Series, Vol. 11447,
  Society of Photo-Optical Instrumentation Engineers (SPIE) Conference Series,
  1144767

\bibitem[{{Gaia Collaboration} {et~al.}(2016){Gaia Collaboration}, {Prusti},
  {de Bruijne}, {Brown}, {Vallenari}, {Babusiaux}, {Bailer-Jones}, {Bastian},
  {Biermann}, {Evans}, {Eyer}, {Jansen}, {Jordi}, {Klioner}, {Lammers},
  {Lindegren}, {Luri}, {Mignard}, {Milligan}, {Panem}, {Poinsignon},
  {Pourbaix}, {Randich}, {Sarri}, {Sartoretti}, {Siddiqui}, {Soubiran},
  {Valette}, {van Leeuwen}, {Walton}, {Aerts}, {Arenou}, {Cropper}, {Drimmel},
  {H{\o}g}, {Katz}, {Lattanzi}, {O'Mullane}, {Grebel}, {Holland}, {Huc},
  {Passot}, {Bramante}, {Cacciari}, {Casta{\~n}eda}, {Chaoul}, {Cheek}, {De
  Angeli}, {Fabricius}, {Guerra}, {Hern{\'a}ndez}, {Jean-Antoine-Piccolo},
  {Masana}, {Messineo}, {Mowlavi}, {Nienartowicz}, {Ord{\'o}{\~n}ez-Blanco},
  {Panuzzo}, {Portell}, {Richards}, {Riello}, {Seabroke}, {Tanga},
  {Th{\'e}venin}, {Torra}, {Els}, {Gracia-Abril}, {Comoretto},
  {Garcia-Reinaldos}, {Lock}, {Mercier}, {Altmann}, {Andrae}, {Astraatmadja},
  {Bellas-Velidis}, {Benson}, {Berthier}, {Blomme}, {Busso}, {Carry},
  {Cellino}, {Clementini}, {Cowell}, {Creevey}, {Cuypers}, {Davidson}, {De
  Ridder}, {de Torres}, {Delchambre}, {Dell'Oro}, {Ducourant}, {Fr{\'e}mat},
  {Garc{\'\i}a-Torres}, {Gosset}, {Halbwachs}, {Hambly}, {Harrison}, {Hauser},
  {Hestroffer}, {Hodgkin}, {Huckle}, {Hutton}, {Jasniewicz}, {Jordan},
  {Kontizas}, {Korn}, {Lanzafame}, {Manteiga}, {Moitinho}, {Muinonen},
  {Osinde}, {Pancino}, {Pauwels}, {Petit}, {Recio-Blanco}, {Robin}, {Sarro},
  {Siopis}, {Smith}, {Smith}, {Sozzetti}, {Thuillot}, {van Reeven}, {Viala},
  {Abbas}, {Abreu Aramburu}, {Accart}, {Aguado}, {Allan}, {Allasia},
  {Altavilla}, {{\'A}lvarez}, {Alves}, {Anderson}, {Andrei}, {Anglada Varela},
  {Antiche}, {Antoja}, {Ant{\'o}n}, {Arcay}, {Atzei}, {Ayache}, {Bach},
  {Baker}, {Balaguer-N{\'u}{\~n}ez}, {Barache}, {Barata}, {Barbier}, {Barblan},
  {Baroni}, {Barrado y Navascu{\'e}s}, {Barros}, {Barstow}, {Becciani},
  {Bellazzini}, {Bellei}, {Bello Garc{\'\i}a}, {Belokurov}, {Bendjoya},
  {Berihuete}, {Bianchi}, {Bienaym{\'e}}, {Billebaud}, {Blagorodnova},
  {Blanco-Cuaresma}, {Boch}, {Bombrun}, {Borrachero}, {Bouquillon}, {Bourda},
  {Bouy}, {Bragaglia}, {Breddels}, {Brouillet}, {Br{\"u}semeister},
  {Bucciarelli}, {Budnik}, {Burgess}, {Burgon}, {Burlacu}, {Busonero}, {Buzzi},
  {Caffau}, {Cambras}, {Campbell}, {Cancelliere}, {Cantat-Gaudin}, {Carlucci},
  {Carrasco}, {Castellani}, {Charlot}, {Charnas}, {Charvet}, {Chassat},
  {Chiavassa}, {Clotet}, {Cocozza}, {Collins}, {Collins}, {Costigan}, {Crifo},
  {Cross}, {Crosta}, {Crowley}, {Dafonte}, {Damerdji}, {Dapergolas}, {David},
  {David}, {De Cat}, {de Felice}, {de Laverny}, {De Luise}, {De March}, {de
  Martino}, {de Souza}, {Debosscher}, {del Pozo}, {Delbo}, {Delgado},
  {Delgado}, {di Marco}, {Di Matteo}, {Diakite}, {Distefano}, {Dolding}, {Dos
  Anjos}, {Drazinos}, {Dur{\'a}n}, {Dzigan}, {Ecale}, {Edvardsson}, {Enke},
  {Erdmann}, {Escolar}, {Espina}, {Evans}, {Eynard Bontemps}, {Fabre},
  {Fabrizio}, {Faigler}, {Falc{\~a}o}, {Farr{\`a}s Casas}, {Faye}, {Federici},
  {Fedorets}, {Fern{\'a}ndez-Hern{\'a}ndez}, {Fernique}, {Fienga}, {Figueras},
  {Filippi}, {Findeisen}, {Fonti}, {Fouesneau}, {Fraile}, {Fraser}, {Fuchs},
  {Furnell}, {Gai}, {Galleti}, {Galluccio}, {Garabato}, {Garc{\'\i}a-Sedano},
  {Gar{\'e}}, {Garofalo}, {Garralda}, {Gavras}, {Gerssen}, {Geyer}, {Gilmore},
  {Girona}, {Giuffrida}, {Gomes}, {Gonz{\'a}lez-Marcos},
  {Gonz{\'a}lez-N{\'u}{\~n}ez}, {Gonz{\'a}lez-Vidal}, {Granvik}, {Guerrier},
  {Guillout}, {Guiraud}, {G{\'u}rpide}, {Guti{\'e}rrez-S{\'a}nchez}, {Guy},
  {Haigron}, {Hatzidimitriou}, {Haywood}, {Heiter}, {Helmi}, {Hobbs},
  {Hofmann}, {Holl}, {Holland}, {Hunt}, {Hypki}, {Icardi}, {Irwin}, {Jevardat
  de Fombelle}, {Jofr{\'e}}, {Jonker}, {Jorissen}, {Julbe}, {Karampelas},
  {Kochoska}, {Kohley}, {Kolenberg}, {Kontizas}, {Koposov}, {Kordopatis},
  {Koubsky}, {Kowalczyk}, {Krone-Martins}, {Kudryashova}, {Kull}, {Bachchan},
  {Lacoste-Seris}, {Lanza}, {Lavigne}, {Le Poncin-Lafitte}, {Lebreton},
  {Lebzelter}, {Leccia}, {Leclerc}, {Lecoeur-Taibi}, {Lemaitre}, {Lenhardt},
  {Leroux}, {Liao}, {Licata}, {Lindstr{\o}m}, {Lister}, {Livanou}, {Lobel},
  {L{\"o}ffler}, {L{\'o}pez}, {Lopez-Lozano}, {Lorenz}, {Loureiro},
  {MacDonald}, {Magalh{\~a}es Fernandes}, {Managau}, {Mann}, {Mantelet},
  {Marchal}, {Marchant}, {Marconi}, {Marie}, {Marinoni}, {Marrese},
  {Marschalk{\'o}}, {Marshall}, {Mart{\'\i}n-Fleitas}, {Martino}, {Mary},
  {Matijevi{\v{c}}}, {Mazeh}, {McMillan}, {Messina}, {Mestre}, {Michalik},
  {Millar}, {Miranda}, {Molina}, {Molinaro}, {Molinaro}, {Moln{\'a}r},
  {Moniez}, {Montegriffo}, {Monteiro}, {Mor}, {Mora}, {Morbidelli}, {Morel},
  {Morgenthaler}, {Morley}, {Morris}, {Mulone}, {Muraveva}, {Musella},
  {Narbonne}, {Nelemans}, {Nicastro}, {Noval}, {Ord{\'e}novic},
  {Ordieres-Mer{\'e}}, {Osborne}, {Pagani}, {Pagano}, {Pailler}, {Palacin},
  {Palaversa}, {Parsons}, {Paulsen}, {Pecoraro}, {Pedrosa}, {Pentik{\"a}inen},
  {Pereira}, {Pichon}, {Piersimoni}, {Pineau}, {Plachy}, {Plum}, {Poujoulet},
  {Pr{\v{s}}a}, {Pulone}, {Ragaini}, {Rago}, {Rambaux}, {Ramos-Lerate},
  {Ranalli}, {Rauw}, {Read}, {Regibo}, {Renk}, {Reyl{\'e}}, {Ribeiro},
  {Rimoldini}, {Ripepi}, {Riva}, {Rixon}, {Roelens}, {Romero-G{\'o}mez},
  {Rowell}, {Royer}, {Rudolph}, {Ruiz-Dern}, {Sadowski}, {Sagrist{\`a}
  Sell{\'e}s}, {Sahlmann}, {Salgado}, {Salguero}, {Sarasso}, {Savietto},
  {Schnorhk}, {Schultheis}, {Sciacca}, {Segol}, {Segovia}, {Segransan},
  {Serpell}, {Shih}, {Smareglia}, {Smart}, {Smith}, {Solano}, {Solitro},
  {Sordo}, {Soria Nieto}, {Souchay}, {Spagna}, {Spoto}, {Stampa}, {Steele},
  {Steidelm{\"u}ller}, {Stephenson}, {Stoev}, {Suess}, {S{\"u}veges}, {Surdej},
  {Szabados}, {Szegedi-Elek}, {Tapiador}, {Taris}, {Tauran}, {Taylor},
  {Teixeira}, {Terrett}, {Tingley}, {Trager}, {Turon}, {Ulla}, {Utrilla},
  {Valentini}, {van Elteren}, {Van Hemelryck}, {van Leeuwen}, {Varadi},
  {Vecchiato}, {Veljanoski}, {Via}, {Vicente}, {Vogt}, {Voss}, {Votruba},
  {Voutsinas}, {Walmsley}, {Weiler}, {Weingrill}, {Werner}, {Wevers},
  {Whitehead}, {Wyrzykowski}, {Yoldas}, {{\v{Z}}erjal}, {Zucker}, {Zurbach},
  {Zwitter}, {Alecu}, {Allen}, {Allende Prieto}, {Amorim},
  {Anglada-Escud{\'e}}, {Arsenijevic}, {Azaz}, {Balm}, {Beck}, {Bernstein},
  {Bigot}, {Bijaoui}, {Blasco}, {Bonfigli}, {Bono}, {Boudreault}, {Bressan},
  {Brown}, {Brunet}, {Bunclark}, {Buonanno}, {Butkevich}, {Carret}, {Carrion},
  {Chemin}, {Ch{\'e}reau}, {Corcione}, {Darmigny}, {de Boer}, {de Teodoro}, {de
  Zeeuw}, {Delle Luche}, {Domingues}, {Dubath}, {Fodor}, {Fr{\'e}zouls},
  {Fries}, {Fustes}, {Fyfe}, {Gallardo}, {Gallegos}, {Gardiol}, {Gebran},
  {Gomboc}, {G{\'o}mez}, {Grux}, {Gueguen}, {Heyrovsky}, {Hoar}, {Iannicola},
  {Isasi Parache}, {Janotto}, {Joliet}, {Jonckheere}, {Keil}, {Kim},
  {Klagyivik}, {Klar}, {Knude}, {Kochukhov}, {Kolka}, {Kos}, {Kutka}, {Lainey},
  {LeBouquin}, {Liu}, {Loreggia}, {Makarov}, {Marseille}, {Martayan},
  {Martinez-Rubi}, {Massart}, {Meynadier}, {Mignot}, {Munari}, {Nguyen},
  {Nordlander}, {Ocvirk}, {O'Flaherty}, {Olias Sanz}, {Ortiz}, {Osorio},
  {Oszkiewicz}, {Ouzounis}, {Palmer}, {Park}, {Pasquato}, {Peltzer}, {Peralta},
  {P{\'e}turaud}, {Pieniluoma}, {Pigozzi}, {Poels}, {Prat}, {Prod'homme},
  {Raison}, {Rebordao}, {Risquez}, {Rocca-Volmerange}, {Rosen}, {Ruiz-Fuertes},
  {Russo}, {Sembay}, {Serraller Vizcaino}, {Short}, {Siebert}, {Silva},
  {Sinachopoulos}, {Slezak}, {Soffel}, {Sosnowska}, {Strai{\v{z}}ys}, {ter
  Linden}, {Terrell}, {Theil}, {Tiede}, {Troisi}, {Tsalmantza}, {Tur},
  {Vaccari}, {Vachier}, {Valles}, {Van Hamme}, {Veltz}, {Virtanen}, {Wallut},
  {Wichmann}, {Wilkinson}, {Ziaeepour}, \& {Zschocke}}]{Gaia2016}
{Gaia Collaboration}, {Prusti}, T., {de Bruijne}, J.~H.~J., {et~al.} 2016,
  \aap, 595, A1

\bibitem[{Gal-Yam(2017)}]{Gal-Yam2017}
Gal-Yam, A. 2017, Observational and Physical Classification of Supernovae, ed.
  A.~W. Alsabti \& P.~Murdin (Cham: Springer International Publishing),
  195--237

\bibitem[{Gehrels {et~al.}(2004)Gehrels, Chincarini, Giommi, Mason, Nousek,
  Wells, White, Barthelmy, Burrows, Cominsky, Hurley, Marshall, Meszaros,
  Roming, Angelini, Barbier, Belloni, Campana, Caraveo, Chester, Citterio,
  Cline, Cropper, Cummings, Dean, Feigelson, Fenimore, Frail, Fruchter,
  Garmire, Gendreau, Ghisellini, Greiner, Hill, Hunsberger, Krimm, Kulkarni,
  Kumar, Lebrun, Lloyd-Ronning, Markwardt, Mattson, Mushotzky, Norris, Osborne,
  Paczynski, Palmer, Park, Parsons, Paul, Rees, Reynolds, Rhoads, Sasseen,
  Schaefer, Short, Smale, Smith, Stella, Tagliaferri, Takahashi, Tashiro,
  Townsley, Tueller, Turner, Vietri, Voges, Ward, Willingale, Zerbi, \&
  Zhang}]{Gehrels_2004}
Gehrels, N., Chincarini, G., Giommi, P., {et~al.} 2004, \apj, 611, 1005

\bibitem[{{Goldberg} {et~al.}(2019){Goldberg}, {Bildsten}, \&
  {Paxton}}]{Goldberg2019}
{Goldberg}, J.~A., {Bildsten}, L., \& {Paxton}, B. 2019, \apj, 879, 3

\bibitem[{{Graham} {et~al.}(2019){Graham}, {Kulkarni}, {Bellm}, {Adams},
  {Barbarino}, {Blagorodnova}, {Bodewits}, {Bolin}, {Brady}, {Cenko}, {Chang},
  {Coughlin}, {De}, {Eadie}, {Farnham}, {Feindt}, {Franckowiak}, {Fremling},
  {Gezari}, {Ghosh}, {Goldstein}, {Golkhou}, {Goobar}, {Ho}, {Huppenkothen},
  {Ivezi{\'c}}, {Jones}, {Juric}, {Kaplan}, {Kasliwal}, {Kelley}, {Kupfer},
  {Lee}, {Lin}, {Lunnan}, {Mahabal}, {Miller}, {Ngeow}, {Nugent}, {Ofek},
  {Prince}, {Rauch}, {van Roestel}, {Schulze}, {Singer}, {Sollerman}, {Taddia},
  {Yan}, {Ye}, {Yu}, {Barlow}, {Bauer}, {Beck}, {Belicki}, {Biswas}, {Brinnel},
  {Brooke}, {Bue}, {Bulla}, {Burruss}, {Connolly}, {Cromer}, {Cunningham},
  {Dekany}, {Delacroix}, {Desai}, {Duev}, {Feeney}, {Flynn}, {Frederick},
  {Gal-Yam}, {Giomi}, {Groom}, {Hacopians}, {Hale}, {Helou}, {Henning},
  {Hover}, {Hillenbrand}, {Howell}, {Hung}, {Imel}, {Ip}, {Jackson}, {Kaspi},
  {Kaye}, {Kowalski}, {Kramer}, {Kuhn}, {Landry}, {Laher}, {Mao}, {Masci},
  {Monkewitz}, {Murphy}, {Nordin}, {Patterson}, {Penprase}, {Porter},
  {Rebbapragada}, {Reiley}, {Riddle}, {Rigault}, {Rodriguez}, {Rusholme}, {van
  Santen}, {Shupe}, {Smith}, {Soumagnac}, {Stein}, {Surace}, {Szkody}, {Terek},
  {Van Sistine}, {van Velzen}, {Vestrand}, {Walters}, {Ward}, {Zhang}, \&
  {Zolkower}}]{Graham2020}
{Graham}, M.~J., {Kulkarni}, S.~R., {Bellm}, E.~C., {et~al.} 2019, \pasp, 131,
  078001

\bibitem[{{Graziani} {et~al.}(2019){Graziani}, {Courtois}, {Lavaux}, {Hoffman},
  {Tully}, {Copin}, \& {Pomar{\`e}de}}]{Graziani2019}
{Graziani}, R., {Courtois}, H.~M., {Lavaux}, G., {et~al.} 2019, \mnras, 488,
  5438

\bibitem[{{Grossan} {et~al.}(1999){Grossan}, {Spillar}, {Tripp}, {Pirzkal},
  {Sutin}, {Johnson}, \& {Barnaby}}]{Grossan1999}
{Grossan}, B., {Spillar}, E., {Tripp}, R., {et~al.} 1999, \aj, 118, 705

\bibitem[{{Guillochon} {et~al.}(2017){Guillochon}, {Parrent}, {Kelley}, \&
  {Margutti}}]{OpenSupernovaCatalog}
{Guillochon}, J., {Parrent}, J., {Kelley}, L.~Z., \& {Margutti}, R. 2017, \apj,
  835, 64

\bibitem[{{Herter} {et~al.}(2008){Herter}, {Henderson}, {Wilson}, {Matthews},
  {Rahmer}, {Bonati}, {Muirhead}, {Adams}, {Lloyd}, {Skrutskie}, {Moon},
  {Parshley}, {Nelson}, {Martinache}, \& {Gull}}]{TSPEC}
{Herter}, T.~L., {Henderson}, C.~P., {Wilson}, J.~C., {et~al.} 2008, in Society
  of Photo-Optical Instrumentation Engineers (SPIE) Conference Series, Vol.
  7014, Ground-based and Airborne Instrumentation for Astronomy II, ed. I.~S.
  {McLean} \& M.~M. {Casali}, 70140X

\bibitem[{{HI4PI Collaboration} {et~al.}(2016){HI4PI Collaboration}, {Ben
  Bekhti}, {Fl{\"o}er}, {Keller}, {Kerp}, {Lenz}, {Winkel}, {Bailin},
  {Calabretta}, {Dedes}, {Ford}, {Gibson}, {Haud}, {Janowiecki}, {Kalberla},
  {Lockman}, {McClure-Griffiths}, {Murphy}, {Nakanishi}, {Pisano}, \&
  {Staveley-Smith}}]{HI4PI2016a}
{HI4PI Collaboration}, {Ben Bekhti}, N., {Fl{\"o}er}, L., {et~al.} 2016, \aap,
  594, A116

\bibitem[{{Horesh} {et~al.}(2020){Horesh}, {Sfaradi}, {Ergon}, {Barbarino},
  {Sollerman}, {Moldon}, {Dobie}, {Schulze}, {P{\'e}rez-Torres}, {Williams},
  {Fremling}, {Gal-Yam}, {Kulkarni}, {O'Brien}, {Lundqvist}, {Murphy},
  {Fender}, {Anand}, {Belicki}, {Bellm}, {Coughlin}, {De}, {Golkhou}, {Graham},
  {Green}, {Hankins}, {Kasliwal}, {Kupfer}, {Laher}, {Masci}, {Miller},
  {Neill}, {Ofek}, {Perrott}, {Porter}, {Reiley}, {Rigault}, {Rodriguez},
  {Rusholme}, {Shupe}, \& {Titterington}}]{horesh_2020}
{Horesh}, A., {Sfaradi}, I., {Ergon}, M., {et~al.} 2020, \apj, 903, 132

\bibitem[{{Horesh} {et~al.}(2013){Horesh}, {Stockdale}, {Fox}, {Frail},
  {Carpenter}, {Kulkarni}, {Ofek}, {Gal-Yam}, {Kasliwal}, {Arcavi}, {Quimby},
  {Cenko}, {Nugent}, {Bloom}, {Law}, {Poznanski}, {Gorbikov}, {Polishook},
  {Yaron}, {Ryder}, {Weiler}, {Bauer}, {Van Dyk}, {Immler}, {Panagia},
  {Pooley}, \& {Kassim}}]{Assaf2013}
{Horesh}, A., {Stockdale}, C., {Fox}, D.~B., {et~al.} 2013, \mnras, 436, 1258

\bibitem[{{Jencson} {et~al.}(2019){Jencson}, {Kasliwal}, {Adams}, {Bond}, {De},
  {Johansson}, {Karambelkar}, {Lau}, {Tinyanont}, {Ryder}, {Cody}, {Masci},
  {Bally}, {Blagorodnova}, {Castell{\'o}n}, {Fremling}, {Gehrz}, {Helou},
  {Kilpatrick}, {Milne}, {Morrell}, {Perley}, {Phillips}, {Smith}, {van Dyk},
  \& {Williams}}]{Jencson2019}
{Jencson}, J.~E., {Kasliwal}, M.~M., {Adams}, S.~M., {et~al.} 2019, \apj, 886,
  40

\bibitem[{{Jerkstrand}(2011)}]{Jerkstrand2011}
{Jerkstrand}, A. 2011, PhD thesis, -

\bibitem[{{Jerkstrand} {et~al.}(2015){Jerkstrand}, {Ergon}, {Smartt},
  {Fransson}, {Sollerman}, {Taubenberger}, {Bersten}, \&
  {Spyromilio}}]{Jerkstrand2015}
{Jerkstrand}, A., {Ergon}, M., {Smartt}, S.~J., {et~al.} 2015, \aap, 573, A12

\bibitem[{{Jerkstrand} {et~al.}(2018){Jerkstrand}, {Ertl}, {Janka},
  {M{\"u}ller}, {Sukhbold}, \& {Woosley}}]{Jerkstrand2018}
{Jerkstrand}, A., {Ertl}, T., {Janka}, H.~T., {et~al.} 2018, \mnras, 475, 277

\bibitem[{{Jerkstrand} {et~al.}(2012){Jerkstrand}, {Fransson}, {Maguire},
  {Smartt}, {Ergon}, \& {Spyromilio}}]{Jerkstrand2012}
{Jerkstrand}, A., {Fransson}, C., {Maguire}, K., {et~al.} 2012, \aap, 546, A28

\bibitem[{{Jerkstrand} {et~al.}(2014){Jerkstrand}, {Smartt}, {Fraser},
  {Fransson}, {Sollerman}, {Taddia}, \& {Kotak}}]{Jerkstrand2014}
{Jerkstrand}, A., {Smartt}, S.~J., {Fraser}, M., {et~al.} 2014, \mnras, 439,
  3694

\bibitem[{{Kaiser} {et~al.}(2002){Kaiser}, {Aussel}, {Burke}, {Boesgaard},
  {Chambers}, {Chun}, {Heasley}, {Hodapp}, {Hunt}, {Jedicke}, {Jewitt},
  {Kudritzki}, {Luppino}, {Maberry}, {Magnier}, {Monet}, {Onaka}, {Pickles},
  {Rhoads}, {Simon}, {Szalay}, {Szapudi}, {Tholen}, {Tonry}, {Waterson}, \&
  {Wick}}]{Kaiser2002}
{Kaiser}, N., {Aussel}, H., {Burke}, B.~E., {et~al.} 2002, in Society of
  Photo-Optical Instrumentation Engineers (SPIE) Conference Series, Vol. 4836,
  Survey and Other Telescope Technologies and Discoveries, ed. J.~A. {Tyson} \&
  S.~{Wolff}, 154--164

\bibitem[{Kamble {et~al.}(2016)Kamble, Margutti, Soderberg, Chakraborti,
  Fransson, Chevalier, Powell, Milisavljevic, Parrent, \&
  Bietenholz}]{Kamble_2016}
Kamble, A., Margutti, R., Soderberg, A.~M., {et~al.} 2016, \apj, 818, 111

\bibitem[{{Kasliwal} {et~al.}(2017){Kasliwal}, {Bally}, {Masci}, {Cody},
  {Bond}, {Jencson}, {Tinyanont}, {Cao}, {Contreras}, {Dykhoff}, {Amodeo},
  {Armus}, {Boyer}, {Cantiello}, {Carlon}, {Cass}, {Cook}, {Corgan}, {Faella},
  {Fox}, {Green}, {Gehrz}, {Helou}, {Hsiao}, {Johansson}, {Khan}, {Lau},
  {Langer}, {Levesque}, {Milne}, {Mohamed}, {Morrell}, {Monson}, {Moore},
  {Ofek}, {O' Sullivan}, {Parthasarathy}, {Perez}, {Perley}, {Phillips},
  {Prince}, {Shenoy}, {Smith}, {Surace}, {Van Dyk}, {Whitelock}, \&
  {Williams}}]{Kasliwal2017}
{Kasliwal}, M.~M., {Bally}, J., {Masci}, F., {et~al.} 2017, \apj, 839, 88

\bibitem[{Kool {et~al.}(2017)Kool, Ryder, Kankare, Mattila, Reynolds, McDermid,
  Pérez-Torres, Herrero-Illana, Schirmer, Efstathiou, Bauer, Kotilainen,
  Väisänen, Baldwin, Romero-Cañizales, \& Alberdi}]{Kool2017}
Kool, E.~C., Ryder, S., Kankare, E., {et~al.} 2017, Monthly Notices of the
  Royal Astronomical Society, 473, 5641

\bibitem[{{Leonard} {et~al.}(2006){Leonard}, {Filippenko}, {Ganeshalingam},
  {Serduke}, {Li}, {Swift}, {Gal-Yam}, {Foley}, {Fox}, {Park}, {Hoffman}, \&
  {Wong}}]{Leonard2006}
{Leonard}, D.~C., {Filippenko}, A.~V., {Ganeshalingam}, M., {et~al.} 2006,
  \nat, 440, 505

\bibitem[{{Li} {et~al.}(2011){Li}, {Chornock}, {Leaman}, {Filippenko},
  {Poznanski}, {Wang}, {Ganeshalingam}, \& {Mannucci}}]{Ratepaper}
{Li}, W., {Chornock}, R., {Leaman}, J., {et~al.} 2011, \mnras, 412, 1473

\bibitem[{{Lourie} {et~al.}(2020){Lourie}, {Baker}, {Burruss}, {Egan},
  {F{\.z}r{\'e}sz}, {Frostig}, {Garcia-Zych}, {Ganciu}, {Haworth},
  {Hinrichsen}, {Kasliwal}, {Karambelkar}, {Malonis}, {Simcoe}, \&
  {Zolkower}}]{Lourie2021}
{Lourie}, N.~P., {Baker}, J.~W., {Burruss}, R.~S., {et~al.} 2020, in Society of
  Photo-Optical Instrumentation Engineers (SPIE) Conference Series, Vol. 11447,
  Society of Photo-Optical Instrumentation Engineers (SPIE) Conference Series,
  114479K

\bibitem[{{Maiolino} {et~al.}(2002){Maiolino}, {Vanzi}, {Mannucci}, {Cresci},
  {Ghinassi}, \& {Della Valle}}]{Maiolino2002}
{Maiolino}, R., {Vanzi}, L., {Mannucci}, F., {et~al.} 2002, \aap, 389, 84

\bibitem[{{Martin} {et~al.}(2018){Martin}, {Fitzgerald}, {McLean}, {Doppmann},
  {Kassis}, {Aliado}, {Canfield}, {Johnson}, {Kress}, {Lanclos}, {Magnone},
  {Sohn}, {Wang}, \& {Weiss}}]{NIRES}
{Martin}, E.~C., {Fitzgerald}, M.~P., {McLean}, I.~S., {et~al.} 2018, in
  Society of Photo-Optical Instrumentation Engineers (SPIE) Conference Series,
  Vol. 10702, Ground-based and Airborne Instrumentation for Astronomy VII, ed.
  C.~J. {Evans}, L.~{Simard}, \& H.~{Takami}, 107020A

\bibitem[{{Martinez} \& {Bersten}(2019)}]{Martinez2019}
{Martinez}, L. \& {Bersten}, M.~C. 2019, \aap, 629, A124

\bibitem[{{Masci} {et~al.}(2019){Masci}, {Laher}, {Rusholme}, {Shupe}, {Groom},
  {Surace}, {Jackson}, {Monkewitz}, {Beck}, {Flynn}, {Terek}, {Landry},
  {Hacopians}, {Desai}, {Howell}, {Brooke}, {Imel}, {Wachter}, {Ye}, {Lin},
  {Cenko}, {Cunningham}, {Rebbapragada}, {Bue}, {Miller}, {Mahabal}, {Bellm},
  {Patterson}, {Juri{\'c}}, {Golkhou}, {Ofek}, {Walters}, {Graham}, {Kasliwal},
  {Dekany}, {Kupfer}, {Burdge}, {Cannella}, {Barlow}, {Van Sistine}, {Giomi},
  {Fremling}, {Blagorodnova}, {Levitan}, {Riddle}, {Smith}, {Helou}, {Prince},
  \& {Kulkarni}}]{Masci2019}
{Masci}, F.~J., {Laher}, R.~R., {Rusholme}, B., {et~al.} 2019, \pasp, 131,
  018003

\bibitem[{{Mattila} {et~al.}(2012){Mattila}, {Dahlen}, {Efstathiou}, {Kankare},
  {Melinder}, {Alonso-Herrero}, {P{\'e}rez-Torres}, {Ryder},
  {V{\"a}is{\"a}nen}, \& {{\"O}stlin}}]{Mattila2012}
{Mattila}, S., {Dahlen}, T., {Efstathiou}, A., {et~al.} 2012, \apj, 756, 111

\bibitem[{{McMullin} {et~al.}(2007){McMullin}, {Waters}, {Schiebel}, {Young},
  \& {Golap}}]{McMullin2007}
{McMullin}, J.~P., {Waters}, B., {Schiebel}, D., {Young}, W., \& {Golap}, K.
  2007, in Astronomical Society of the Pacific Conference Series, Vol. 376,
  Astronomical Data Analysis Software and Systems XVI, ed. R.~A. {Shaw},
  F.~{Hill}, \& D.~J. {Bell}, 127

\bibitem[{{Miluzio, M.} {et~al.}(2013){Miluzio, M.}, {Cappellaro, E.},
  {Botticella, M. T.}, {Cresci, G.}, {Greggio, L.}, {Mannucci, F.}, {Benetti,
  S.}, {Bufano, F.}, {Elias-Rosa, N.}, {Pastorello, A.}, {Turatto, M.}, \&
  {Zampieri, L.}}]{Miluzio2013}
{Miluzio, M.}, {Cappellaro, E.}, {Botticella, M. T.}, {et~al.} 2013, A\&A, 554,
  A127

\bibitem[{{Moore} \& {Kasliwal}(2019)}]{Moore2020}
{Moore}, A. \& {Kasliwal}, M. 2019, Nature Astronomy, 3, 109

\bibitem[{{Morozova} {et~al.}(2015){Morozova}, {Piro}, {Renzo}, {Ott},
  {Clausen}, {Couch}, {Ellis}, \& {Roberts}}]{Morozova2015}
{Morozova}, V., {Piro}, A.~L., {Renzo}, M., {et~al.} 2015, \apj, 814, 63

\bibitem[{{Morozova} {et~al.}(2017){Morozova}, {Piro}, \&
  {Valenti}}]{Morozova2017}
{Morozova}, V., {Piro}, A.~L., \& {Valenti}, S. 2017, \apj, 838, 28

\bibitem[{{Morozova} {et~al.}(2018){Morozova}, {Piro}, \&
  {Valenti}}]{Morozova2018}
{Morozova}, V., {Piro}, A.~L., \& {Valenti}, S. 2018, \apj, 858, 15

\bibitem[{{Nagao} {et~al.}(2017){Nagao}, {Maeda}, \& {Tanaka}}]{Nagao2017}
{Nagao}, T., {Maeda}, K., \& {Tanaka}, M. 2017, \apj, 847, 111

\bibitem[{{Nagao} {et~al.}(2018){Nagao}, {Maeda}, \& {Tanaka}}]{Nagao2018}
{Nagao}, T., {Maeda}, K., \& {Tanaka}, M. 2018, \apj, 861, 1

\bibitem[{{Nakamura} {et~al.}(2016){Nakamura}, {Horiuchi}, {Tanaka}, {Hayama},
  {Takiwaki}, \& {Kotake}}]{Nakamura2016}
{Nakamura}, K., {Horiuchi}, S., {Tanaka}, M., {et~al.} 2016, \mnras, 461, 3296

\bibitem[{{Oke} {et~al.}(1995){Oke}, {Cohen}, {Carr}, {Cromer}, {Dingizian},
  {Harris}, {Labrecque}, {Lucinio}, {Schaal}, {Epps}, \& {Miller}}]{LRIS}
{Oke}, J.~B., {Cohen}, J.~G., {Carr}, M., {et~al.} 1995, \pasp, 107, 375

\bibitem[{Oke \& Gunn(1982)}]{Oke_1982}
Oke, J.~B. \& Gunn, J.~E. 1982, \pasp, 94, 586

\bibitem[{{Patterson} {et~al.}(2019){Patterson}, {Bellm}, {Rusholme}, {Masci},
  {Juric}, {Krughoff}, {Golkhou}, {Graham}, {Kulkarni}, {Helou}, \& {Zwicky
  Transient Facility Collaboration}}]{Patterson2019}
{Patterson}, M.~T., {Bellm}, E.~C., {Rusholme}, B., {et~al.} 2019, \pasp, 131,
  018001

\bibitem[{{Perley}(2019)}]{Perley2019}
{Perley}, D.~A. 2019, \pasp, 131, 084503

\bibitem[{Rabinak \& Waxman(2011)}]{Rabinak_2011}
Rabinak, I. \& Waxman, E. 2011, \apj, 728, 63

\bibitem[{Rayner {et~al.}(2003)Rayner, Toomey, Onaka, Denault, Stahlberger,
  Vacca, Cushing, \& Wang}]{IRTF}
Rayner, J.~T., Toomey, D.~W., Onaka, P.~M., {et~al.} 2003, \pasp, 115, 362

\bibitem[{{Richardson} {et~al.}(2014){Richardson}, {Jenkins}, {Wright}, \&
  {Maddox}}]{Richardson2014}
{Richardson}, D., {Jenkins}, Robert~L., I., {Wright}, J., \& {Maddox}, L. 2014,
  \aj, 147, 118

\bibitem[{{Rigault} {et~al.}(2019){Rigault}, {Neill}, {Blagorodnova}, {Dugas},
  {Feeney}, {Walters}, {Brinnel}, {Copin}, {Fremling}, {Nordin}, \&
  {Sollerman}}]{Rigault2019}
{Rigault}, M., {Neill}, J.~D., {Blagorodnova}, N., {et~al.} 2019, \aap, 627,
  A115

\bibitem[{{Roming} {et~al.}(2005){Roming}, {Kennedy}, {Mason}, {Nousek}, {Ahr},
  {Bingham}, {Broos}, {Carter}, {Hancock}, {Huckle}, {Hunsberger}, {Kawakami},
  {Killough}, {Koch}, {McLelland}, {Smith}, {Smith}, {Soto}, {Boyd},
  {Breeveld}, {Holland}, {Ivanushkina}, {Pryzby}, {Still}, \&
  {Stock}}]{Roming2005a}
{Roming}, P. W.~A., {Kennedy}, T.~E., {Mason}, K.~O., {et~al.} 2005, \ssr, 120,
  95

\bibitem[{Sanders {et~al.}(2015)Sanders, Soderberg, Gezari, Betancourt,
  Chornock, Berger, Foley, Challis, Drout, Kirshner, Lunnan, Marion, Margutti,
  McKinnon, Milisavljevic, Narayan, Rest, Kankare, Mattila, Smartt, Huber,
  Burgett, Draper, Hodapp, Kaiser, Kudritzki, Magnier, Metcalfe, Morgan, Price,
  Tonry, Wainscoat, \& Waters}]{Sanders_2015}
Sanders, N.~E., Soderberg, A.~M., Gezari, S., {et~al.} 2015, \apj, 799, 208

\bibitem[{Sapir \& Waxman(2017)}]{Sapir_2017}
Sapir, N. \& Waxman, E. 2017, \apj, 838, 130

\bibitem[{{Schlafly} \& {Finkbeiner}(2011)}]{Schlafly2011}
{Schlafly}, E.~F. \& {Finkbeiner}, D.~P. 2011, \apj, 737, 103

\bibitem[{{Shaya} {et~al.}(2017){Shaya}, {Tully}, {Hoffman}, \&
  {Pomar{\`e}de}}]{Shaya2017}
{Shaya}, E.~J., {Tully}, R.~B., {Hoffman}, Y., \& {Pomar{\`e}de}, D. 2017,
  \apj, 850, 207

\bibitem[{Shivvers {et~al.}(2015)Shivvers, Groh, Mauerhan, Fox, Leonard, \&
  Filippenko}]{Shivvers_2015}
Shivvers, I., Groh, J.~H., Mauerhan, J.~C., {et~al.} 2015, \apj, 806, 213

\bibitem[{{Simcoe} {et~al.}(2019){Simcoe}, {F{\H{u}}r{\'e}sz}, {Sullivan},
  {Hellickson}, {Malonis}, {Kasliwal}, {Shectman}, {Kollmeier}, \&
  {Moore}}]{Simcoe2019}
{Simcoe}, R.~A., {F{\H{u}}r{\'e}sz}, G., {Sullivan}, P.~W., {et~al.} 2019, \aj,
  157, 46

\bibitem[{Sivanandam {et~al.}(2018)Sivanandam, Chapman, Simard, Hickson, Venn,
  Thibault, Sawicki, Muzzin, Erickson, Abraham, Akiyama, Andersen, Bradley,
  Carlberg, Chen, Correia, Davidge, Ellison, El-Sankary, Fahlman, Lamb,
  Lardière, Lemoine-Busserolle, Moon, Murray, Peck, Shafai, Sivo, Veran, \&
  Yee}]{Gemini}
Sivanandam, S., Chapman, S., Simard, L., {et~al.} 2018, in Ground-based and
  Airborne Instrumentation for Astronomy VII, ed. C.~J. Evans, L.~Simard, \&
  H.~Takami, Vol. 10702, International Society for Optics and Photonics (SPIE),
  456 -- 467

\bibitem[{{Smartt}(2009)}]{Smartt2009}
{Smartt}, S.~J. 2009, \araa, 47, 63

\bibitem[{{Smartt}(2015)}]{Smartt2015}
{Smartt}, S.~J. 2015, \pasa, 32, e016

\bibitem[{Smith(2014)}]{radioreview}
Smith, N. 2014, Annual Review of Astronomy and Astrophysics, 52, 487–528

\bibitem[{{Soderberg} {et~al.}(2012){Soderberg}, {Margutti}, {Zauderer},
  {Krauss}, {Katz}, {Chomiuk}, {Dittmann}, {Nakar}, {Sakamoto}, {Kawai},
  {Hurley}, {Barthelmy}, {Toizumi}, {Morii}, {Chevalier}, {Gurwell},
  {Petitpas}, {Rupen}, {Alexander}, {Levesque}, {Fransson}, {Brunthaler},
  {Bietenholz}, {Chugai}, {Grindlay}, {Copete}, {Connaughton}, {Briggs},
  {Meegan}, {von Kienlin}, {Zhang}, {Rau}, {Golenetskii}, {Mazets}, \&
  {Cline}}]{soderberg_2012}
{Soderberg}, A.~M., {Margutti}, R., {Zauderer}, B.~A., {et~al.} 2012, \apj,
  752, 78

\bibitem[{{Sorce} {et~al.}(2014){Sorce}, {Tully}, {Courtois}, {Jarrett},
  {Neill}, \& {Shaya}}]{2014distance}
{Sorce}, J.~G., {Tully}, R.~B., {Courtois}, H.~M., {et~al.} 2014, \mnras, 444,
  527

\bibitem[{{Spiro} {et~al.}(2014){Spiro}, {Pastorello}, {Pumo}, {Zampieri},
  {Turatto}, {Smartt}, {Benetti}, {Cappellaro}, {Valenti}, {Agnoletto},
  {Altavilla}, {Aoki}, {Brocato}, {Corsini}, {Di Cianno}, {Elias-Rosa},
  {Hamuy}, {Enya}, {Fiaschi}, {Folatelli}, {Desidera}, {Harutyunyan}, {Howell},
  {Kawka}, {Kobayashi}, {Leibundgut}, {Minezaki}, {Navasardyan}, {Nomoto},
  {Mattila}, {Pietrinferni}, {Pignata}, {Raimondo}, {Salvo}, {Schmidt},
  {Sollerman}, {Spyromilio}, {Taubenberger}, {Valentini}, {Vennes}, \&
  {Yoshii}}]{Spiro2014}
{Spiro}, S., {Pastorello}, A., {Pumo}, M.~L., {et~al.} 2014, \mnras, 439, 2873

\bibitem[{{Sukhbold} {et~al.}(2016){Sukhbold}, {Ertl}, {Woosley}, {Brown}, \&
  {Janka}}]{Sukhbold2016}
{Sukhbold}, T., {Ertl}, T., {Woosley}, S.~E., {Brown}, J.~M., \& {Janka}, H.~T.
  2016, \apj, 821, 38

\bibitem[{{Tinyanont} {et~al.}(2019){Tinyanont}, {Millar-Blanchaer}, {Nilsson},
  {Mawet}, {Knutson}, {Kataria}, {Vasisht}, {Henderson}, {Matthews}, {Serabyn},
  {Milburn}, {Hale}, {Smith}, {Vissapragada}, {Santos}, {Kekas}, \&
  {Escuti}}]{Tinyanont2019a}
{Tinyanont}, S., {Millar-Blanchaer}, M.~A., {Nilsson}, R., {et~al.} 2019,
  \pasp, 131, 025001

\bibitem[{{Tomasella } {et~al.}(2014){Tomasella }, {Benetti}, {Cappellaro},
  {Pastorello}, {Turatto}, {Barbon}, {Elias-Rosa}, {Harutyunyan}, {Ochner},
  {Tartaglia}, \& {Valenti}}]{Asiago}
{Tomasella }, L., {Benetti}, S., {Cappellaro}, E., {et~al.} 2014, Astronomische
  Nachrichten, 335, 841

\bibitem[{{Tully} {et~al.}(2016){Tully}, {Courtois}, \&
  {Sorce}}]{Tullydistance}
{Tully}, R.~B., {Courtois}, H.~M., \& {Sorce}, J.~G. 2016, \aj, 152, 50

\bibitem[{{Tully} \& {Fisher}(1977)}]{TullyFischerPaper}
{Tully}, R.~B. \& {Fisher}, J.~R. 1977, \aap, 500, 105

\bibitem[{{Tully} \& {Fisher}(1988)}]{Bbanddistance}
{Tully}, R.~B. \& {Fisher}, J.~R. 1988, {Catalog of Nearby Galaxies}

\bibitem[{{Uomoto}(1986)}]{Uomoto1986}
{Uomoto}, A. 1986, \apjl, 310, L35

\bibitem[{{Utrobin} \& {Chugai}(2009)}]{Utrobin2009}
{Utrobin}, V.~P. \& {Chugai}, N.~N. 2009, \aap, 506, 829

\bibitem[{{Utrobin} \& {Chugai}(2015)}]{Utrobin2015}
{Utrobin}, V.~P. \& {Chugai}, N.~N. 2015, \aap, 575, A100

\bibitem[{{Utrobin} \& {Chugai}(2017)}]{Utrobin2017}
{Utrobin}, V.~P. \& {Chugai}, N.~N. 2017, \mnras, 472, 5004

\bibitem[{{Vacca} {et~al.}(2003){Vacca}, {Cushing}, \& {Rayner}}]{Vacca2003}
{Vacca}, W.~D., {Cushing}, M.~C., \& {Rayner}, J.~T. 2003, \pasp, 115, 389

\bibitem[{{Wang} \& {Wheeler}(2008)}]{Wang2008}
{Wang}, L. \& {Wheeler}, J.~C. 2008, \araa, 46, 433

\bibitem[{Weiler {et~al.}(2002)Weiler, Panagia, Montes, \&
  Sramek}]{Weiler_2002}
Weiler, K.~W., Panagia, N., Montes, M.~J., \& Sramek, R.~A. 2002, Annual Review
  of Astronomy and Astrophysics, 40, 387

\bibitem[{{Woosley} \& {Heger}(2007)}]{Woosley2007}
{Woosley}, S.~E. \& {Heger}, A. 2007, \physrep, 442, 269

\bibitem[{{Yadav} {et~al.}(2014){Yadav}, {Ray}, {Chakraborti}, {Stockdale},
  {Chandra}, {Smith}, {Roy}, {Bose}, {Dwarkadas}, {Sutaria}, \&
  {Pooley}}]{yadav_2014}
{Yadav}, N., {Ray}, A., {Chakraborti}, S., {et~al.} 2014, \apj, 782, 30

\bibitem[{{Zackay} {et~al.}(2016){Zackay}, {Ofek}, \& {Gal-Yam}}]{Zackay2016}
{Zackay}, B., {Ofek}, E.~O., \& {Gal-Yam}, A. 2016, \apj, 830, 27

\end{thebibliography}

\end{document}